\documentclass[english,aps,tightenlines,showpacs, showkeys, notitlepage]{revtex4-2}
\usepackage[T1]{fontenc}
\usepackage[latin9]{inputenc}
\usepackage{color}
\usepackage{babel}
\usepackage{amstext}
\usepackage{graphicx}
\usepackage{setspace}
\usepackage{esint}
\usepackage[unicode=true,pdfusetitle,
 bookmarks=true,bookmarksnumbered=false,bookmarksopen=false,
 breaklinks=false,pdfborder={0 0 0},pdfborderstyle={},backref=false,colorlinks=true]
 {hyperref}
\begin{document}
\title{Anisotropic Born--Infeld-$f(R)$ Cosmologies}
\author{Salih Kibaro\u{g}lu$^{1}$}
\email{salihkibaroglu@maltepe.edu.tr}

\date{\today}
\begin{abstract}
In this paper, we investigate anisotropic cosmological solutions within
the framework of Born-Infeld-$f(R)$ gravity, a modification of general
relativity that incorporates higher-order curvature invariants. Specifically,
we focus on the analysis of Bianchi type-I solutions, which provide
a valuable framework for studying the evolution of spatial anisotropies
in cosmology. By introducing a reconstruction scheme for this model,
we explore the behavior of these solutions, analyze various cosmic
evolution scenarios, including Starobinsky inflation, de Sitter-like
expansion, power-law dynamics, and Big Rip-like evolution. Additionally,
we examine the exponential and power-law behavior of the auxiliary
metric function to illustrate its impact on the dynamics of cosmic
evolution. Our analysis elucidates how the modifications introduced
by Born-Infeld-$f(R)$ gravity influence anisotropic dynamics, offering
new perspectives on the late-time behavior of the universe and potential
future singularities.
\end{abstract}
\affiliation{$^{1}$Maltepe University, Faculty of Engineering and Natural Sciences,
34857, Istanbul, Türkiye}
\maketitle

\section{Introduction}

The formulation of general relativity (GR) by Albert Einstein in 1915
marked a pivotal moment in the history of theoretical physics, fundamentally
altering our understanding of gravity and spacetime. This theory has
been remarkably successful in explaining a wide range of gravitational
phenomena, from the precession of Mercury's orbit to the expansion
of the universe. However, despite its successes, general relativity
faces significant challenges, particularly at the quantum level and
in describing the very early universe. These limitations have spurred
the development of modified theories of gravity, such as $f\left(R\right)$
theories \citep{nojiri2011unified,nojiri2007introduction,Nojiri:2017ncd},
which extend general relativity by introducing a more general dependence
on the Ricci scalar $R$.

Born-Infeld (BI) gravity is a nonlinear generalization of Einstein's
general relativity, inspired by the BI theory of electromagnetism
\citep{born1934foundations}, which was originally introduced to remove
singularities in the electric field of point charges. This concept
was later adapted to gravity, leading to the BI gravitational action,
which features a square root determinant structure and introduces
higher-order curvature terms. The resulting approach, in principle,
avoids singularities like those found in black hole solutions \citep{deser1998born}.
However, the inclusion of higher-order curvature terms in this framework
can lead to the emergence of ghost modes, which represent unphysical
degrees of freedom that can destabilize the theory at the quantum
level. To address the ghost problem, the Palatini formulation of BI
gravity, inspired by Eddington's theory \citep{eddington1923mathematical},
offers a promising solution by avoiding the higher-order derivatives
in the equations of motion that typically lead to ghost modes \citep{vollick2004palatini,vollick2005born,Banados2010eddington}.
This approach is known as \textquotedbl Eddington-inspired Born-Infeld
gravity\textquotedbl{} (EiBI). As a result, the implications of EiBI
gravitational theory have been thoroughly explored across various
fields, including cosmology \citep{Avelino:2012BIcosmologies,Escamilla-Rivera:2012Tensor,Cho:2012Universe,Scargill:2012Cosmology,Kruglov:2013Modified,Yang:2013Linear,Du:2014Large,Kim:2014Origin},
astrophysics \citep{Harko:2015Wormhole,Avelino:2012BIcosmologies},
black hole physics \citep{Olmo:2014GeonicBH}, and wormhole physics
\citep{Lobo:2014Microscopic,Harko:2015Wormhole}. Moreover, efforts
have been made to extend EiBI gravitational theory to higher curvature
scenarios by incorporating an $f\left(R\right)$ Lagrangian into the
original framework \citep{makarenko2014born}. The combination of
EiBI gravity with $f\left(R\right)$ modifications has led to a rich
framework known as BI-$f\left(R\right)$ gravity, allowing for a more
comprehensive description of gravitational phenomena which holds promise
for resolving some of the outstanding problems in modern cosmology
and gravitational theory \citep{makarenko2014born,Makarenko:2014nca,odintsov2014born,Elizalde:2016vsd,Kibaroglu:2023BIdS,Kibaroglu:2023CosmBI,Kibaroglu:2024UnimodularBI}
(for a detailed review, see \citep{jimenez2018born}).

Cosmological models frequently assume isotropy, implying that the
universe appears identical in all directions. However, recent observations,
such as those from the Wilkinson Microwave Anisotropy Probe (WMAP)
dataset \citep{WMAP:2003,WMAP:2006,WMAP:2008}, highlight the need
to extend beyond the standard isotropic and homogeneous cosmological
framework. The inflationary paradigm, supported by observational evidence,
demonstrates a remarkable ability to drive an initially anisotropic
and inhomogeneous universe toward the isotropic and homogeneous FLRW
geometry observed today. Nevertheless, achieving a comprehensive understanding
of cosmic evolution requires relaxing the assumptions of perfect isotropy
and homogeneity to include spatial anisotropy and inhomogeneity, along
with an investigation of their evolution into the near-isotropic state
observed at present. As a first step, Bianchi-type cosmological models
provide a valuable framework, forming a broad and nearly complete
class of relativistic cosmological models that preserve homogeneity
but allow for deviations from isotropy. Over the past few decades,
Bianchi cosmologies have gained renewed interest in observational
cosmology, particularly due to WMAP data \citep{WMAP:2003}, which
suggest that the standard cosmological model with a positive cosmological
constant closely resembles the Bianchi morphology \citep{Jaffe:2005Evidence,Jaffe:2006Bianchi,Jaffe:2006Fast,Campanelli:2006Ellipsoidal,Campanelli:2007Cosmic}.
These findings indicate that the universe may have retained a slightly
anisotropic spatial geometry, a result at odds with the predictions
of conventional inflationary models \citep{guth1981inflationary,linde1982new,linde1983chaotic,linde1991axions,linde1994hybrid}.
Recently, a wide range of Bianchi cosmological models have been investigated
from various perspectives, providing deeper insights into the nature
of primordial fluctuations and the large-scale structure of the universe.
These studies offer a promising avenue for addressing open questions
in cosmology (see \citep{Collins:1980Exact,Ellis:2006TheBianchiModels,Gumrukcuoglu:2007Inflationary,Akarsu:2010BianchiTypeIII,Muller:2018Anisotropic,Amirhashchi:2020Constraining,Kumar:2021Anisotropic,Costantini:2022AReconstruction,Nojiri:2022Formalizing,Parnovsky:2023TheBigBang}
for further details).

Integrating anisotropic cosmologies with BI gravity offers a novel
approach to addressing these complexities, potentially leading to
new predictions and a deeper understanding of cosmological evolution
\citep{Vollick:2003AnisotropicBI,Rodrigues:2008Evolution,Harko:2014Bianchi}.
From this background, we direct our attention to the properties of
the BI-$f\left(R\right)$ theory, with a particular emphasis on its
implications for anisotropic cosmic evolution, including its potential
to provide refined models of early universe dynamics and deviations
from isotropy in the large-scale structure.

The paper is organized as follows: after giving a brief discussion
of the BI-$f\left(R\right)$ theory in Section \ref{sec:Born-Infeld--gravity},
we introduce the solution of the field equation in the context of
homogeneous and anisotropic universe model with Bianchi Type I metric.
In Section \ref{sec:Reconstruction-Method-Together}, we discuss the
applications of BI-$f\left(R\right)$ gravity to specific cosmological
scenarios by considering the reconstruction method in the context
of axisymmetric solution of the anisotropic metric. The paper ends
with some conclusive remarks and discussions in the final section.

\section{\label{sec:Born-Infeld--gravity}Born-Infeld-$f\left(R\right)$ gravity}

Let us briefly review the standard BI-$f\left(R\right)$ theory \citep{makarenko2014born}
(see also \citep{Makarenko:2014nca,odintsov2014born}). In this theory,
the original EiBI gravity theory is combined with an additional $f\left(R\right)$
that depends on the Ricci scalar $R=g^{\mu\nu}R_{\mu\nu}\left(\Gamma\right)$.
To avoid any ghost instabilities, the theory is formulated within
the Palatini formalism (for more detail see \citep{Olmo:2011uz,olmo2012open}),
in which the metric $g_{\mu\nu}$ and the connection $\Gamma_{\beta\gamma}^{\alpha}$
are treated as independent variables. The action for this theory is
given by

\begin{eqnarray}
S & = & \frac{2}{\kappa^{2}\epsilon}\int\text{d}^{4}x\left[\sqrt{-|g_{\mu\nu}+\epsilon R_{\mu\nu}|}-\lambda\sqrt{-g}\right]+\frac{\alpha}{2\kappa^{2}}\int\text{d}^{4}x\left[\sqrt{-g}f\left(R\right)\right]+S_{m},\label{eq: action_f(R)}
\end{eqnarray}
where the first term represents the standard BI gravitational Lagrangian
and the second term is an additional function of the Ricci scalar.
$S_{m}$ is the matter action which is only coupled to the metric
$g_{\mu\nu}$ as dictated by the equivalence principle. The parameter
$\kappa$ is a constant with inverse dimensions to that of a cosmological
constant and $\lambda$ is a dimensionless constant. Note that $g$
is the determinant of the metric. Throughout this paper, we will use
Planck units $8\pi G=1$ and set the speed of light to $c=1$. 

In the limit as $\epsilon\rightarrow0$, the BI Lagrangian reduces
to the standard GR term, and the action simplifies to an $f\left(R\right)$
theory. Alternatively, if we take the limit $\alpha\rightarrow0$,
we retrieve the BI theory. When we take both limits, GR is naturally
recovered. In this formulation, we assume vanishing torsion and a
symmetric Ricci tensor. For the sake of simplicity, we assume the
conditions $\kappa=1$ and $\alpha=1$ throughout this paper.

The field equations can be derived from Eq.(\ref{eq: action_f(R)})
through independent variation with respect to the metric and connection.
The variation of this action with respect to the metric tensor leads
to a modified metric field equations for the standard BI gravitational
model,
\begin{equation}
\frac{\sqrt{-q}}{\sqrt{-g}}\left(q^{-1}\right)^{\mu\nu}-\lambda g^{\mu\nu}+\frac{\epsilon}{2}g^{\mu\nu}f\left(R\right)-\epsilon f_{R}R^{\mu\nu}=-\epsilon T^{\mu\nu},\label{eq: eom_g-1}
\end{equation}
where $f_{R}$ is the derivative of $f\left(R\right)$ with respect
to the Ricci scalar, and $T^{\mu\nu}$ is the standard energy-momentum
tensor. In the above equations, we have used the notation,
\begin{equation}
q_{\mu\nu}=g_{\mu\nu}+\epsilon R_{\mu\nu},\label{eq: q_f(R)-1}
\end{equation}
and we denoted the inverse of $q_{\mu\nu}$ by $\left(q^{-1}\right)^{\mu\nu}$
and $q$ represents the determinant of $q_{\mu\nu}$. Similarly, the
connection $\Gamma_{\beta\gamma}^{\alpha}$ variation has the form;

\begin{equation}
\nabla_{\lambda}\left(\sqrt{-q}q^{\mu\nu}+\sqrt{-g}f_{R}g^{\mu\nu}\right)=0,\label{eq: eom_conn-1}
\end{equation}
where the covariant derivative is taken with respect to the independent
connection which is defined for a scalar field $\phi$ as
\begin{equation}
\nabla_{\mu}\phi=\partial_{\mu}\phi-\Gamma_{\mu\alpha}^{\alpha}\phi.
\end{equation}

Furthermore, it is well established in the literature on Palatini
$f\left(R\right)$ theories that the independent connection can be
expressed in terms of an auxiliary metric $h_{\mu\nu}$, which is
conformal to $g_{\mu\nu}$ (for more detail, see \citep{Olmo:2011uz}).
Assuming that $q_{\mu\nu}$ is conformally proportional to the metric
tensor as 
\begin{equation}
q_{\mu\nu}=k\left(t\right)g_{\mu\nu},\label{eq: conf_rel}
\end{equation}
thus Eq.(\ref{eq: eom_conn-1}) becomes,
\begin{equation}
\nabla_{\nu}\left[\sqrt{-u}\left(u^{-1}\right)^{\mu\nu}\right]=0.\label{eq: eom_u}
\end{equation}
In this case, we have an auxiliary metric defined as $u_{\mu\nu}=\left[k\left(t\right)+f_{R}\right]g_{\mu\nu}$
and $\left(u^{-1}\right)^{\mu\nu}$ represents the inverse representation
of $u_{\mu\nu}$. Eq.(\ref{eq: eom_u}) tells us that the connection
can be defined by this auxiliary metric as
\begin{equation}
\Gamma_{\mu\nu}^{\rho}=\frac{1}{2}\left(u^{-1}\right)^{\rho\sigma}\left(u_{\sigma\nu,\mu}+u_{\mu\sigma,\nu}-u_{\mu\nu,\sigma}\right).
\end{equation}
By considering the conformal relation between $q_{\mu\nu}$ and $g_{\mu\nu}$
in Eq.(\ref{eq: conf_rel}), we can easily say that the Ricci tensor
must also be proportional to the metric tensor. Thus one can write
the relationship between the Ricci tensor and $g_{\mu\nu}$, as
\begin{equation}
R_{\mu\nu}=\frac{1}{\epsilon}\left[k\left(t\right)-1\right]g_{\mu\nu}.\label{eq: Ricci_g}
\end{equation}

\subsection{Eximining the anisotropic cosmology}

We now proceed to analyze a homogeneous and anisotropic space-time
described by the Bianchi type-I metric, which represents the simplest
generalization of the spatially flat FLRW metric. This class of metrics
is characterized by its spatial homogeneity but allows for directional
anisotropy, making it a suitable framework for studying deviations
from isotropy in the universe. The Bianchi type-I metric, expressed
in the form:

\begin{equation}
\text{d}s^{2}=-\text{d}t^{2}+A\left(t\right)^{2}\text{d}x^{2}+B\left(t\right)^{2}\text{d}y^{2}+C\left(t\right)^{2}\text{d}z^{2},\label{eq: metric_B1}
\end{equation}
where $A\left(t\right)$, $B\left(t\right)$ and $C\left(t\right)$
are the scale factors associated with the $x$, $y$ and $z$ axes,
respectively. It is worth noting that if all three scale factors are
equal one obtains flat FLRW metric. We can define the average expansion
scale factor as $a\left(t\right)=\left(ABC\right)^{1/3}$ for this
model. This model is spatially flat, which corresponds to the well-known
fact that our Universe is spatially flat or almost spatially flat. 

Additionally, the anisotropic nature of the expansion can be characterized
by introducing the directional Hubble parameters and defining the
mean Hubble parameter for the overall expansion. The directional Hubble
parameters for different directions defined along the principal $x$,
$y$, $z$ axes;
\begin{equation}
H_{x}\left(t\right)=\frac{\dot{A}}{A},\,\,\,\,\,\,\,\,\,\,\,\,\,H_{y}\left(t\right)=\frac{\dot{B}}{B},\,\,\,\,\,\,\,\,\,\,\,\,\,H_{z}\left(t\right)=\frac{\dot{C}}{C}.
\end{equation}
Here, the mean Hubble parameter becomes $H\left(t\right)=\left(1/3\right)\left(H_{x}+H_{y}+H_{z}\right)$
and an additional parameter $H_{u}\left(t\right)=\dot{u}/u$ is defined
for the scalar parameter $u\left(t\right)$. One can also define the
volume of the universe as $V=ABC$. Furthermore, the anisotropy parameter
of the expansion is defined as

\begin{equation}
\Delta=\frac{1}{3}\sum_{i=1}^{3}\left(\frac{H_{i}-H}{H}\right)^{2},
\end{equation}
where $i=1,2,3$ and $H_{i}$ denote the directional Hubble parameters
corresponding to the $x$, $y$ and $z$ axes, respectively. Here
$\Delta=0$ condition indicates a state of isotropic expansion.

From this background, we can define the auxiliary metric as
\begin{equation}
u_{\mu\nu}=u\left(t\right)\text{diag}\left(-1,A\left(t\right)^{2},B\left(t\right)^{2},C\left(t\right)^{2}\right),\label{eq: metric_U}
\end{equation}
where $u\left(t\right)=k\left(t\right)+f_{R}$. According to Eq.(\ref{eq: Ricci_g}),
we can define the Ricci tensor as 
\begin{eqnarray}
R_{\mu\nu} & = & r\left(t\right)g_{\mu\nu}.\label{eq: Ricci_metric_relation}
\end{eqnarray}
In addition, it can also be shown that the function $r\left(t\right)$
can be written as
\begin{equation}
r\left(t\right)=\frac{1}{\epsilon}\left(u\left(t\right)-f_{R}-1\right).\label{eq: r(t)_def}
\end{equation}

Now, one can find the $r\left(t\right)$ function by taking into account
of $\left\{ t,t\right\} $, $\left\{ x,x\right\} $, $\left\{ y,y\right\} $,
$\left\{ z,z\right\} $ components as follows, respectively
\begin{equation}
r\left(t\right)=\frac{\ddot{A}}{A}+\frac{\ddot{B}}{B}+\frac{\ddot{C}}{C}+\frac{1}{2}H_{x}H_{u}+\frac{1}{2}H_{y}H_{u}+\frac{1}{2}H_{z}H_{u}+\frac{3}{2}\dot{H}_{u},\label{eq: r_1}
\end{equation}

\begin{equation}
r\left(t\right)=\frac{\ddot{A}}{A}+\frac{\ddot{u}}{2u}+H_{x}H_{y}+H_{x}H_{z}+\frac{3}{2}H_{x}H_{u}+\frac{1}{2}H_{y}H_{u}+\frac{1}{2}H_{z}H_{u},\label{eq: r_2}
\end{equation}

\begin{equation}
r\left(t\right)=\frac{\ddot{B}}{B}+\frac{\ddot{u}}{2u}+H_{x}H_{y}+H_{y}H_{z}+\frac{1}{2}H_{x}H_{u}+\frac{3}{2}H_{y}H_{u}+\frac{1}{2}H_{z}H_{u},\label{eq: r_3}
\end{equation}

\begin{equation}
r\left(t\right)=\frac{\ddot{C}}{C}+\frac{\ddot{u}}{2u}+H_{x}H_{z}+H_{y}H_{z}+\frac{1}{2}H_{x}H_{u}+\frac{1}{2}H_{y}H_{u}+\frac{3}{2}H_{z}H_{u},\label{eq: r_4}
\end{equation}
where an overdot on a field variable denotes differentiation with
respect to time $t$. Within this framework, by combining these equations,
an alternative expression for $r\left(t\right)$ can be derived in
the following form:

\begin{equation}
r\left(t\right)=\frac{3}{4}H_{u}^{2}+H_{u}\left(H_{x}+H_{y}+H_{z}\right)+H_{x}H_{y}+H_{x}H_{z}+H_{y}H_{z},
\end{equation}
Moreover, we can additionally express:

\begin{equation}
\frac{\ddot{B}}{B}+\frac{\ddot{C}}{C}+\frac{\ddot{u}}{u}-H_{x}H_{u}-\frac{3}{2}H_{u}^{2}-H_{x}H_{y}-H_{x}H_{z}=0,
\end{equation}

\begin{equation}
\frac{\ddot{A}}{A}+\frac{\ddot{C}}{C}+\frac{\ddot{u}}{u}-H_{y}H_{u}-\frac{3}{2}H_{u}^{2}-H_{x}H_{y}-H_{y}H_{z}=0,
\end{equation}

\begin{equation}
\frac{\ddot{A}}{A}+\frac{\ddot{B}}{B}+\frac{\ddot{u}}{u}-H_{z}H_{u}-\frac{3}{2}H_{u}^{2}-H_{x}H_{z}-H_{y}H_{z}=0.
\end{equation}

\section{Reconstruction Method Together with Axisymmetric solution\label{sec:Reconstruction-Method-Together}}

In this section, we will explore a reconstruction method (for related
approaches, see \citep{Nojiri:2006Modified,Cognola:2007String,Cognola:2008AClass,Elizalde:2008Reconstructing,Nojiri:2009CosmologicalReconstruction})
that possesses the intriguing capability of realizing any specified
evolution with particular scale factors, enabling us to explicitly
derive the BI-$f\left(R\right)$ model that corresponds to the given
cosmological evolution. 

The standard Bianchi type-I metric given in the previous section results
in highly intricate field equations (\ref{eq: Ricci_metric_relation})
together with Eqs.(\ref{eq: r_1})-(\ref{eq: r_4}), making the application
of the reconstruction method significantly challenging. To simplify
the mathematical framework for the provided model, this study adopts
the axisymmetric Bianchi type-I metric, which reduces the complexity
of the equations while still capturing essential features of anisotropic
cosmology. This approach facilitates a more tractable analysis and
enables the derivation of viable cosmological solutions within the
chosen theoretical framework. For this purpose, we consider $A\left(t\right)=B\left(t\right)$
in the metric (\ref{eq: metric_B1}) as a specific subclass of the
Bianchi type-I model that exhibits axisymmetry around the $z$-axis
as described by the metric in Cartesian coordinates;

\begin{equation}
\text{d}s^{2}=-\text{d}t^{2}+A\left(t\right)^{2}\left(\text{d}x^{2}+\text{d}y^{2}\right)+C\left(t\right)^{2}\text{d}z^{2}.\label{eq: metric_B1_axisymmetric}
\end{equation}
This metric, often referred to as the most general plane-symmetric
line element \citep{Taub:1951Empty}, is constructed by selecting
the $xy$-plane as the plane of symmetry. The specific form of the
Bianchi-I metric selected above is motivated by physical sources exhibiting
azimuthal symmetry, as discussed in references \citep{Barrow:2006Cosmologies,Berera:2004TheEccentric,Campanelli:2006Ellipsoidal}.
This spacetime configuration is also commonly referred to in the literature
as the Eccentric or the Ellipsoidal Universe \citep{Berera:2004TheEccentric,Campanelli:2009AModel}.
Additionally, if we set $A\left(t\right)=C\left(t\right)$ in the
metric (\ref{eq: metric_B1_axisymmetric}), without loss of generality
by rescaling the $x$, $y$ and $z$, coordinates, the metric simplifies
to its corresponding isotropic FLRW form.

\subsection{Inflationary period}

To investigate a viable inflationary scenario, we focus on the Starobinsky
inflation model \citep{Starobinsky:1980ANew,Barrow:1988Inflation,Odintsov:2015SingularInflationary},
which has gained significant attention due to its compelling inflationary
phenomenology and strong agreement with observational data, such as
those provided by the Planck mission \citep{Planck:2020Planck}. In
this context, we consider the evolution of the Hubble parameter as
a function of cosmic time, which characterizes the inflationary expansion
and governs the dynamics of the early universe, given explicitly by:

\begin{equation}
H\left(t\right)=H_{0}-\frac{M^{2}}{6}\left(t-t_{i}\right),\label{eq: Hubble_starobinsky}
\end{equation}
with $H_{0}$, $M$, and $t_{i}$ are arbitrary constants. Actually
$t_{i}$ represents the beginning of inflation, which can be considered
as the horizon crossing of the large scale mode ($\sim0.05\mathrm{Mpc}^{-1}$),
and moreover, $H_{0}$ sets the inflationary energy scale at its onset.
The slow roll parameter, defined by $\epsilon_{\mathrm{1}}=-\dot{H}/H^{2}$,
corresponding to the above Hubble parameter comes as, 
\begin{equation}
\epsilon_{\mathrm{1}}=\frac{M^{2}}{6\left\{ H_{0}^{2}-\frac{M^{2}}{3}\left(t-t_{i}\right)\right\} }.\label{e1}
\end{equation}
Clearly, for $M<H_{0}$, the slow roll parameter is less than unity
at $t=t_{i}$ which indicates an accelerated phase of the universe.
Moreover, Eq.(\ref{e1}) also indicates that $\epsilon_{\mathrm{1}}$
is an increasing function of time, and eventually reaches to unity
at 
\begin{equation}
t_{\mathrm{f}}=t_{i}+\frac{3H_{0}}{M^{2}}\left(H_{0}^{2}-\frac{M^{2}}{6}\right),\label{exit time}
\end{equation}
which describes the end of inflation. Thereby unlike to the de-Sitter
case, the Starobinsky inflation (where the Hubble parameter follows
Eq.(\ref{eq: Hubble_starobinsky})) provides an exit of inflation.
Furthermore the inflationary indices in the Starobinsky inflation,
particularly the spectral index of scalar perturbation ($n_{s}$)
and the tensor-to-scalar ratio ($r$) have the well known expressions:
\begin{equation}
n_{s}=1-\frac{2}{N}~~~~~~~\mathrm{and}~~~~~~~r=\frac{12}{N^{2}},
\end{equation}
where $N$ is the total e-folding number of inflationary period. For
$N\approx60$, such observable indices are in good agreement with
the Planck data \citep{Planck:2020Planck}, which makes the Starobinsky
inflation a viable one. In this particular model, since the study
of this model is done for early times, we can approximate this scale
factor as follows,
\begin{equation}
a\left(t\right)\simeq a_{0}e^{H_{0}t+\frac{M^{2}}{6}tt_{i}}.
\end{equation}

This is the general concept of the Starobinsky inflation model. In
order to apply this concept to our model let us define the scale factors
as follows;

\begin{equation}
A\left(t\right)=A_{0}e^{H_{A0}t+\frac{M_{A}^{2}}{6}tt_{i}},\,\,\,\,\,\,\,\,\,\,\,\,\,\,\,C\left(t\right)=C_{0}e^{H_{C0}t+\frac{M_{C}^{2}}{6}tt_{i}}.\label{eq: a_c_def_starobinsky}
\end{equation}

From this background, considering Eq.(\ref{eq: Ricci_metric_relation})
by taking into account of $\left\{ t,t\right\} $, $\left\{ x,x\right\} $,
$\left\{ z,z\right\} $ terms we find the following three equations,
\begin{eqnarray}
r\left(t\right) & = & \frac{3}{2}\dot{H}_{u}+\frac{H_{u}}{6}\left[\left(M_{A}^{2}+\frac{M_{C}^{2}}{2}\right)t_{i}+3\left(2H_{A0}+H_{C0}\right)\right]\nonumber \\
 &  & +\frac{1}{18}\left(M_{A}^{4}+\frac{M_{C}^{4}}{2}\right)t_{i}^{2}+\frac{1}{3}\left(2H_{A0}M_{A}^{2}+H_{C0}M_{C}^{2}\right)t_{i}+\left(2H_{A0}^{2}+H_{C0}^{2}\right),\label{eq: r_starob_tt}
\end{eqnarray}

\begin{eqnarray}
r\left(t\right) & = & \frac{\ddot{u}}{2u}+\frac{H_{u}}{12}\left[\left(4M_{A}^{2}+M_{C}^{2}\right)t_{i}+6\left(4H_{A0}+H_{C}\right)\right]+\frac{1}{36}\left(2M_{A}^{4}+M_{A}^{2}M_{C}^{2}\right)t_{i}^{2}\nonumber \\
 &  & +\frac{1}{6}\left[\left(4M_{A}^{2}+M_{C}^{2}\right)H_{A0}+M_{A}^{2}H_{C0}\right]t_{i}+\left(2H_{A0}^{2}+H_{A0}H_{C0}\right),\label{eq: r_starob_xx}
\end{eqnarray}

\begin{eqnarray}
r\left(t\right) & = & \frac{\ddot{u}}{2u}+\frac{H_{u}}{12}\left[\left(2M_{A}^{2}+3M_{C}^{2}\right)t_{i}+6\left(2H_{A0}+3H_{C}\right)\right]+\frac{1}{36}\left(M_{C}^{4}+2M_{A}^{2}M_{C}^{2}\right)t_{i}^{2}\nonumber \\
 &  & +\frac{1}{3}\left[\left(M_{A}^{2}+M_{C}^{2}\right)H_{C0}+M_{C}^{2}H_{A0}\right]t_{i}+\left(H_{C0}^{2}+2H_{A0}H_{C0}\right).\label{eq: r_starob_zz}
\end{eqnarray}
and subtracting Eq.(\ref{eq: r_starob_tt}) from Eq.(\ref{eq: r_starob_xx})
and (\ref{eq: r_starob_zz}) we get

\begin{equation}
-\frac{1}{6}H_{u}\left(M_{A}^{2}t_{i}+6H_{A0}\right)-\frac{1}{36}\left(M_{C}^{2}t_{i}+6H_{C0}\right)\left[\left(M_{A}^{2}-M_{C}^{2}\right)t_{i}+6\left(H_{A0}-H_{C0}\right)\right]-\frac{3}{2}H_{u}^{2}+\frac{\ddot{u}}{u}=0,\label{eq: a_0_1-2}
\end{equation}

\begin{equation}
-\frac{1}{6}H_{u}\left(M_{C}^{2}t_{i}+6H_{C0}\right)+\frac{1}{18}\left(M_{A}^{2}t_{i}+6H_{A0}\right)\left[\left(M_{A}^{2}-M_{C}^{2}\right)t_{i}+6\left(H_{A0}-H_{C0}\right)\right]-\frac{3}{2}H_{u}^{2}+\frac{\ddot{u}}{u}=0,\label{eq: c_0-1-2}
\end{equation}
then solving Eq.(\ref{eq: c_0-1-2}) with respect to $u\left(t\right)$
we obtain

\begin{equation}
u\left(t\right)=\left[\frac{\mathcal{W}^{2}e^{-\frac{\left(M_{A}^{2}t_{i}+6H_{A0}\right)t}{6}}}{C_{1}e^{-\frac{\mathcal{W}t}{6}}-C_{2}}\right]^{2}.\label{eq: u_sol_Starob}
\end{equation}
where we introduce 
\begin{equation}
\mathcal{W}=\left(2M_{A}^{2}-M_{C}^{2}\right)t_{i}+6\left(2H_{A0}-H_{C0}\right)
\end{equation}
for simplicty. Moreover by using Eq.(\ref{eq: r(t)_def}), we assume
the $f\left(R\right)$ curvature function has the form of $f\left(R\right)=\Phi\left(t\right)R$.
Therefore the auxiliary function can be found as 
\begin{eqnarray}
\Phi\left(t\right) & = & \frac{1}{36\left(C_{1}e^{-\frac{2}{3}\mathcal{W}t}-C_{2}\right)}\Biggl\{-8\epsilon C_{1}C_{2}\left[\left(M_{A}^{2}-M_{C}^{2}\right)t_{i}+6\left(H_{A0}-H_{C0}\right)\right]\left[\left(M_{A}^{2}-\frac{M_{C}^{2}}{4}\right)t_{i}+3\left(H_{A0}-\frac{H_{C0}}{2}\right)\right]e^{-\frac{2}{3}\mathcal{W}t}\nonumber \\
 &  & -4\epsilon C_{1}^{2}\left(M_{A}^{2}t_{i}+6H_{A0}\right)\left[\left(M_{A}^{2}-\frac{M_{C}^{2}}{4}\right)t_{i}+3\left(H_{A0}-\frac{H_{C0}}{2}\right)\right]e^{-\frac{4}{3}\mathcal{W}t}\nonumber \\
 &  & 64\mathcal{W}^{2}e^{-\frac{1}{3}\left(M_{A}^{2}t_{i}+6H_{A0}\right)t}+\epsilon C_{2}^{2}\left(M_{C}^{2}t_{i}+6H_{C0}\right)\left[\left(M_{A}^{2}-M_{C}^{2}\right)t_{i}+6\left(H_{A0}-H_{C0}\right)\right]\Biggr\}\label{eq: Phi_sol_starob}
\end{eqnarray}

\begin{figure}
\includegraphics[width=8cm]{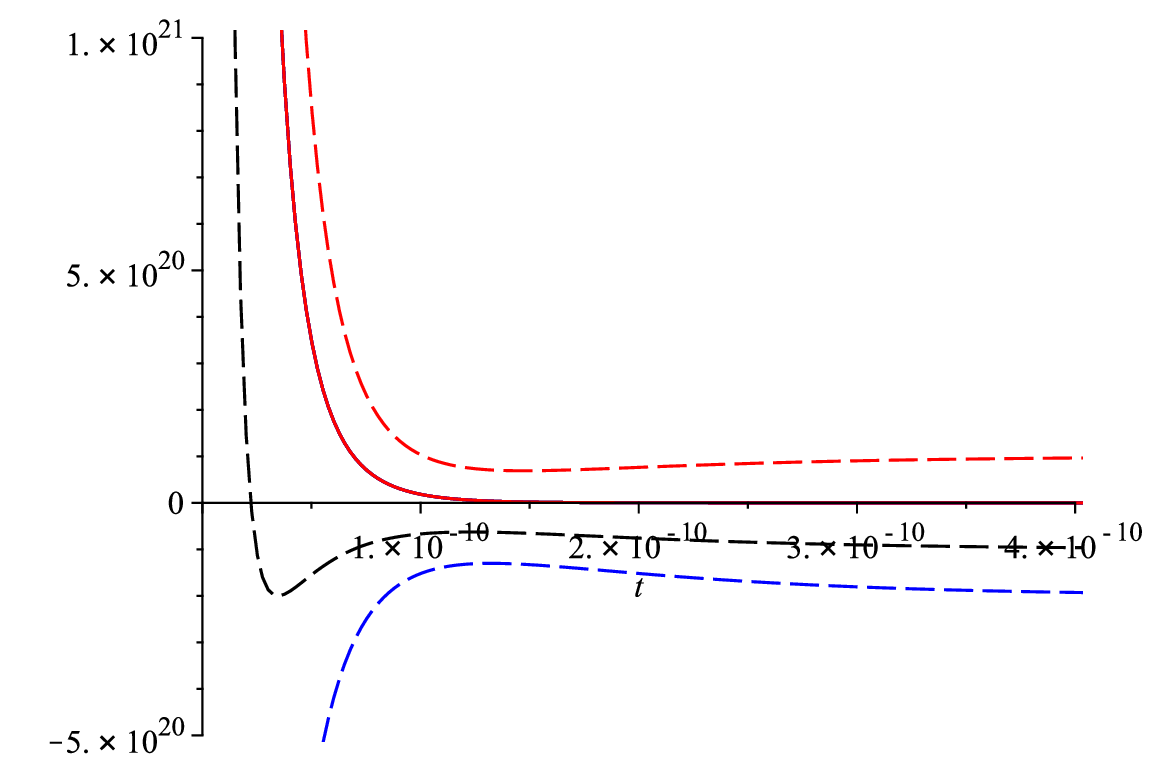}\includegraphics[width=8cm]{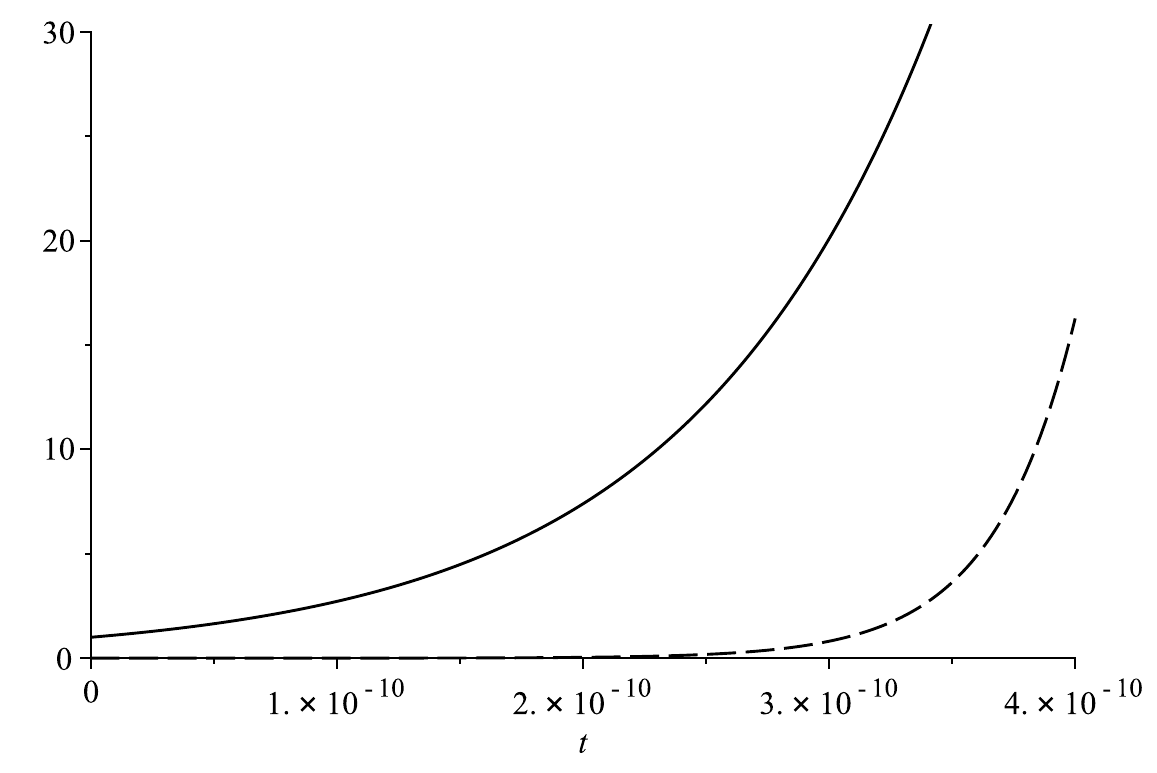}

\caption{Regarding the reconstruction of Starobinsky inflation, the figure
on the left presents $u\left(t\right)$ (solid line) as defined in
Eq.(\ref{eq: u_sol_Starob}) and $\Phi\left(t\right)$ (dashed line)
in Eq.(\ref{eq: Phi_sol_starob}). The figure on the right illustrates
$A\left(t\right)$ (solid line) and $C\left(t\right)$ (dashed line)
in Eq.(\ref{eq: a_c_def_starobinsky}) with parameters $a_{0}=1$,
$H_{A0}=10^{10}$, $M_{A}=10^{-5}$,$c_{0}=10^{-4}$, $H_{C0}=3\times10^{10}$,
$M_{A}=3\times10^{-5}$, $C_{1}=1$ and $C_{2}=2$. For the left plot,
note that $\epsilon=1$ (black line), $\epsilon=2$ (blue line), $\epsilon=-1$
(red line).\label{fig: Starobinsky}}
\end{figure}

In Fig. \ref{fig: Starobinsky}, we plot the functions $u\left(t\right),$$\Phi\left(t\right)$
and the scale factors at early times, by choosing $t_{i}=10^{-35}\text{sec}$.
It is important to note that time is measured in seconds; thus, the
Hubble rate is expressed in $\text{sec}^{-1}$ and the present time
corresponds to $t_{p}\sim10^{17}\text{sec}$.

\subsection{de Sitter like solution}

The de Sitter universe is a solution to Einstein's field equations
and this model is significant in the study of cosmological evolution,
especially in the context of dark energy and the accelerated expansion
of the universe. In this model the corresponding scale factor exponentially
evolves with respect to the cosmic time. Inspired by this framework,
let's consider our scale factors in the following forms;
\begin{equation}
A\left(t\right)=e^{h_{1}t},\,\,\,\,\,\,\,\,\,\,\,\,\,\,\,C\left(t\right)=e^{h_{2}t}.\label{eq: a_c_def_deSitter}
\end{equation}
 where $h_{1}$ and $h_{2}$ are constants. Therefore the auxiliary
metric looks like 
\begin{equation}
u_{\mu\nu}=u\left(t\right)\text{diag}\left(-1,e^{2h_{1}t},e^{2h_{1}t},e^{2h_{2}t}\right).
\end{equation}

From this background, considering Eq.(\ref{eq: Ricci_metric_relation})
by taking into account of $\left\{ t,t\right\} $, $\left\{ x,x\right\} $,
$\left\{ z,z\right\} $ terms we find the following three equations,
\begin{equation}
r\left(t\right)=2h_{1}^{2}+h_{2}^{2}+h_{1}H_{u}+\frac{h_{2}}{2}H_{u}+\frac{3}{2}\dot{H}_{u},\label{eq: r_aks_1-1}
\end{equation}

\begin{equation}
r\left(t\right)=2h_{1}^{2}+h_{1}h_{2}+2h_{1}H_{u}+\frac{h_{2}}{2}H_{u}+\frac{\ddot{u}}{2u},\label{eq: r_aks_2-1}
\end{equation}

\begin{equation}
r\left(t\right)=h_{2}^{2}+2h_{1}h_{2}+h_{1}H_{u}+\frac{3h_{2}}{2}H_{u}+\frac{\ddot{u}}{2u},\label{eq: r_aks_3-1}
\end{equation}
and subtracting Eq.(\ref{eq: r_aks_1-1}) from Eq.(\ref{eq: r_aks_2-1})
and (\ref{eq: r_aks_3-1}) we get

\begin{equation}
h_{2}^{2}-h_{1}h_{2}-h_{1}H_{u}-\frac{3}{2}H_{u}^{2}+\frac{\ddot{u}}{u}=0,\label{eq: a_0_1}
\end{equation}

\begin{equation}
2h_{1}^{2}-2h_{1}h_{2}-h_{2}H_{u}-\frac{3}{2}H_{u}^{2}+\frac{\ddot{u}}{u}=0,\label{eq: c_0-1}
\end{equation}
then solving Eq.(\ref{eq: c_0-1}) with respect to $u\left(t\right)$
we obtain

\begin{equation}
u\left(t\right)=\left[\frac{2\left(2h_{1}-h_{2}\right)}{e^{h_{1}t}\left(C_{1}e^{-\left(2h_{1}-h_{2}\right)t}-C_{2}\right)}\right]^{2}.\label{eq: u_sol_deSitter}
\end{equation}

Moreover by using Eq.(\ref{eq: r(t)_def}), we assume the $f\left(R\right)$
curvature function has the form of $f\left(R\right)=\Phi\left(t\right)R$.
Therefore the auxiliary function can be found as 
\begin{equation}
\Phi\left(t\right)=\frac{-4\epsilon C_{1}\left(h_{1}-\frac{h_{2}}{4}\right)e^{-\left(2h_{1}-h_{2}\right)t}\left[C_{1}h_{1}e^{-\left(2h_{1}-h_{2}\right)t}+2C_{2}\left(h_{1}-h_{2}\right)\right]+\epsilon C_{2}^{2}h_{2}\left(h_{1}-h_{2}\right)+4\left(2h_{1}-h_{2}\right)^{2}e^{-2h_{1}t}}{\left(C_{1}e^{-\left(2h_{1}-h_{2}\right)t}-C_{2}\right)^{2}}-1.\label{eq: Phi_sol_desitter}
\end{equation}

\begin{figure}
\includegraphics[width=8cm]{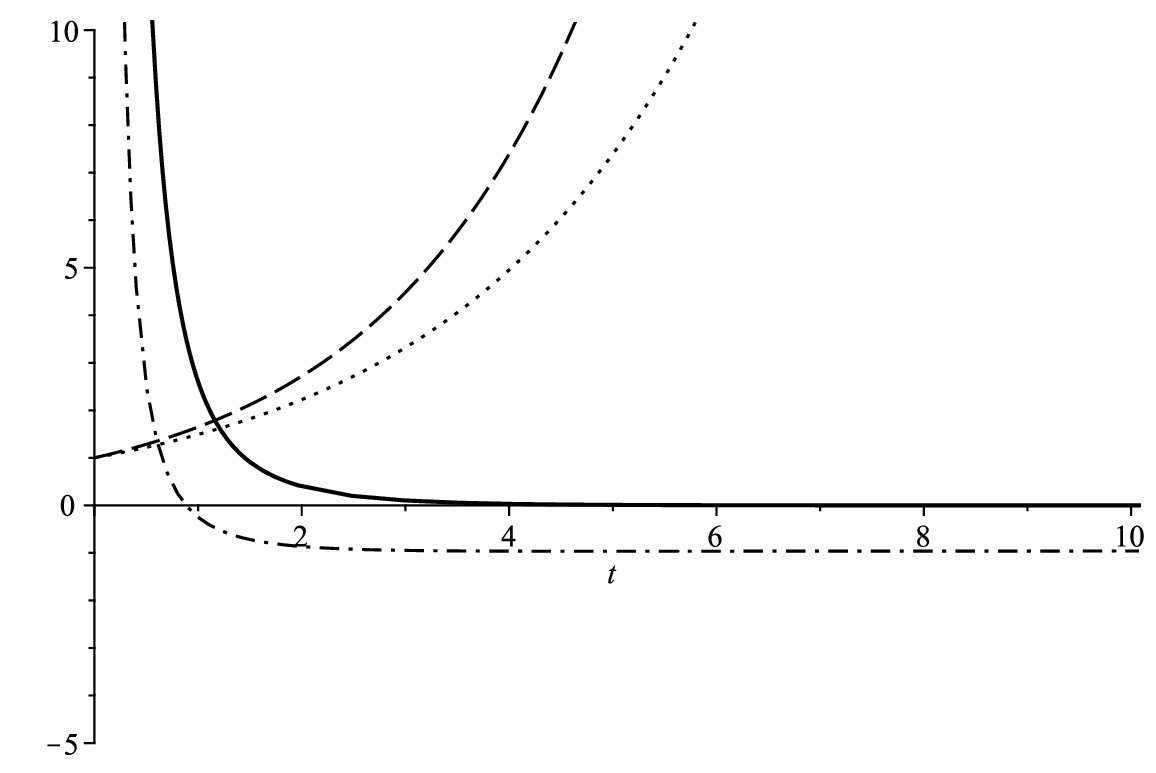}\includegraphics[width=8cm]{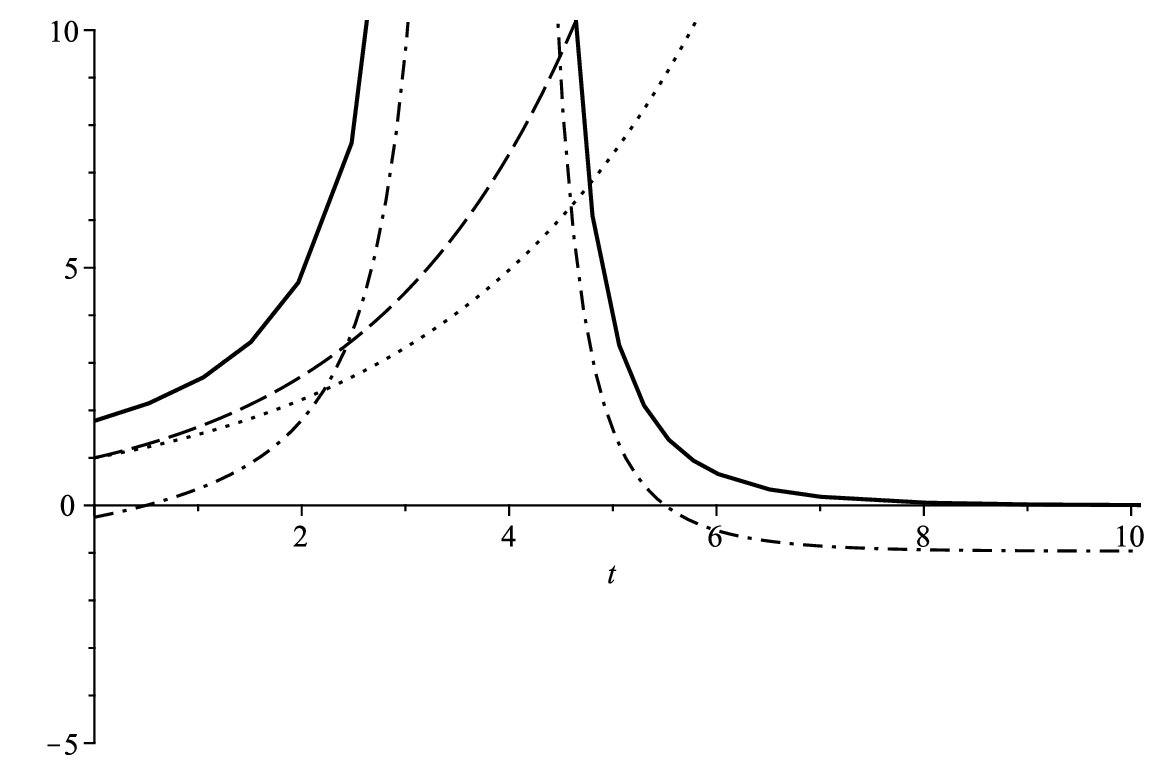}

\caption{Concerning the reconstruction of the de Sitter solution, these figures
show $u\left(t\right)$ (solid line) in Eq.(\ref{eq: u_sol_deSitter}),
$\Phi\left(t\right)$ (dashdot line) in Eq.(\ref{eq: Phi_sol_desitter}),
$A\left(t\right)$ (dashed line) and $C\left(t\right)$ (dotted line)
in Eq.(\ref{eq: a_c_def_deSitter}) with the parameters $h_{1}=0.5$,
$h_{2}=0.4$, $C_{1}=1$ and $C_{2}=1$ for left one and the parameters
$h_{1}=0.5$, $h_{2}=0.4$, $C_{1}=1$ and $C_{2}=0.1$ for the right
one.\label{fig: deSitter}}
\end{figure}

\subsection{Power-law cosmology}

This cosmological scenario refers to a class of solutions to the Friedmann
equations where the scale factor $a\left(t\right)$ evolves as a power
of time like $a\left(t\right)\propto t^{n}$ where $n$ is a constant
that depends on the dominant component of the universe's energy density.
This type of evolution is characteristic of universes dominated by
different types of matter or energy, such as radiation and matter.
Solving the Friedmann equations for a radiation-dominated universe
yields: $a\left(t\right)\propto t^{1/2}$. This leads to the equation
of state (EoS) function $w=1/3$, where $w$ is the ratio of pressure
to energy density. For a matter-dominated universe, the solution to
the Friedmann equations is: $a\left(t\right)\propto t^{2/3}$ ($w=0$).
This describes the expansion of the universe during the matter-dominated
era. 

Let us examine the following case which the scale factors obey,

\begin{equation}
A\left(t\right)=A_{0}\left(\frac{t}{t_{0}}\right)^{h_{1}},\,\,\,\,\,\,\,\,\,\,\,\,\,\,\,C\left(t\right)=C_{0}\left(\frac{t}{t_{0}}\right)^{h_{2}},\label{eq: a_c_def_power}
\end{equation}
where $h_{1}$ and $h_{2}$ are constants, $t_{0}$ is present age
of the universe and $A_{0}$ and $C_{0}$ are the present value of
$A\left(t\right)$ and $C\left(t\right)$, respectively. Therefore
the auxiliary metric looks like 
\begin{equation}
u_{\mu\nu}=u\left(t\right)\text{diag}\left(-1,A_{0}^{2}\left(\frac{t}{t_{0}}\right)^{2h_{1}},A_{0}^{2}\left(\frac{t}{t_{0}}\right)^{2h_{1}},C_{0}^{2}\left(\frac{t}{t_{0}}\right)^{2h_{1}}\right).
\end{equation}

From this background, considering Eq.(\ref{eq: Ricci_metric_relation})
by taking into account of $\left\{ t,t\right\} $, $\left\{ x,x\right\} $,
$\left\{ z,z\right\} $ terms we find the following three equations,
\begin{equation}
r\left(t\right)=\frac{2h_{1}^{2}-2h_{1}-h_{2}+h_{2}^{2}}{t^{2}}+\frac{h_{1}}{t}H_{u}+\frac{h_{2}}{2t}H_{u}+\frac{3}{2}\dot{H}_{u},\label{eq: r_aks_1-1-1}
\end{equation}

\begin{equation}
r\left(t\right)=\frac{2h_{1}^{2}-h_{1}+h_{1}h_{2}}{t^{2}}+\frac{2h_{1}}{t}H_{u}+\frac{h_{2}}{2t}H_{u}+\frac{\ddot{u}}{2u},\label{eq: r_aks_2-1-1}
\end{equation}

\begin{equation}
r\left(t\right)=\frac{2h_{1}h_{2}-h_{2}+h_{2}^{2}}{t^{2}}+\frac{h_{1}}{t}H_{u}+\frac{3h_{2}}{2t}H_{u}+\frac{\ddot{u}}{2u},\label{eq: r_aks_3-1-1}
\end{equation}
and some combination of these equations satisfy

\begin{equation}
\frac{-h_{1}-h_{1}h_{2}-h_{2}+h_{2}^{2}}{t^{2}}-\frac{h_{1}}{t}H_{u}+\frac{3}{2}H_{u}^{2}+\frac{\ddot{u}}{u}=0,\label{eq: a_0_1-1}
\end{equation}

\begin{equation}
\frac{2h_{1}^{2}-h_{1}-2h_{1}h_{2}}{t^{2}}-\frac{h_{2}}{t}H_{u}+\frac{3}{2}H_{u}^{2}+\frac{\ddot{u}}{u}=0,\label{eq: c_0-1-1}
\end{equation}
then solving Eq.(\ref{eq: c_0-1-1}) with respect to $u\left(t\right)$
we obtain

\begin{equation}
u\left(t\right)=\left[\frac{2\left(2h_{1}-h_{2}-1\right)}{t^{h_{1}}\left(C_{1}t^{-\left(2h_{1}-h_{2}-1\right)}-C_{2}\right)}\right]^{2}.\label{eq: u_sol_power}
\end{equation}
Subsequently, using Eq.(\ref{eq: r(t)_def}), one can find the exact
expression for $f\left(R\right)=\Phi\left(t\right)R$ as

\begin{eqnarray}
\Phi\left(t\right) & = & \frac{8}{\mathcal{M}^{4}}\left(h_{1}-\frac{h_{2}}{2}-\frac{1}{2}\right)\Biggl(-\frac{3}{2}\epsilon C_{1}^{3}C_{2}\left(h_{1}-1\right)t^{2h_{1}+3h_{2}+1}+\frac{3}{2}\epsilon C_{1}^{2}C_{2}^{2}\left(h_{2}-1\right)t^{4h_{1}+2h_{2}}\nonumber \\
 &  & +\epsilon C_{1}C_{2}^{3}\left(h_{1}-h_{2}\right)t^{6h_{1}+h_{2}-1}+\epsilon C_{1}^{4}\left(h_{1}-\frac{h_{2}}{4}-\frac{3}{4}\right)t^{4h_{2}+2}-2\frac{\mathcal{M}^{4}}{\mathcal{K}^{2}}\left(h_{1}-\frac{h_{2}}{2}-\frac{1}{2}\right)\Biggr)-1,\label{eq: Phi_sol_power}
\end{eqnarray}
where we have used the following expressions 
\begin{equation}
\mathcal{K}\left(t\right)=C_{1}t^{-2h_{1}+h_{2}+1}-C_{2},\,\,\,\,\,\,\,\,\,\,\,\,\,\,\,\,\,\,\,\,\,\mathcal{M}\left(t\right)=C_{1}t^{h_{2}+1}-C_{2}t^{2h_{1}}.\label{eq: def_A_and_B-1}
\end{equation}
It is easy to see that at the condition $2h_{1}-h_{2}=1$ leads to
$f\left(R\right)$ goes to zero therefore the theory reduces the well-known
BI theory. 

\begin{figure}
\includegraphics[width=8cm]{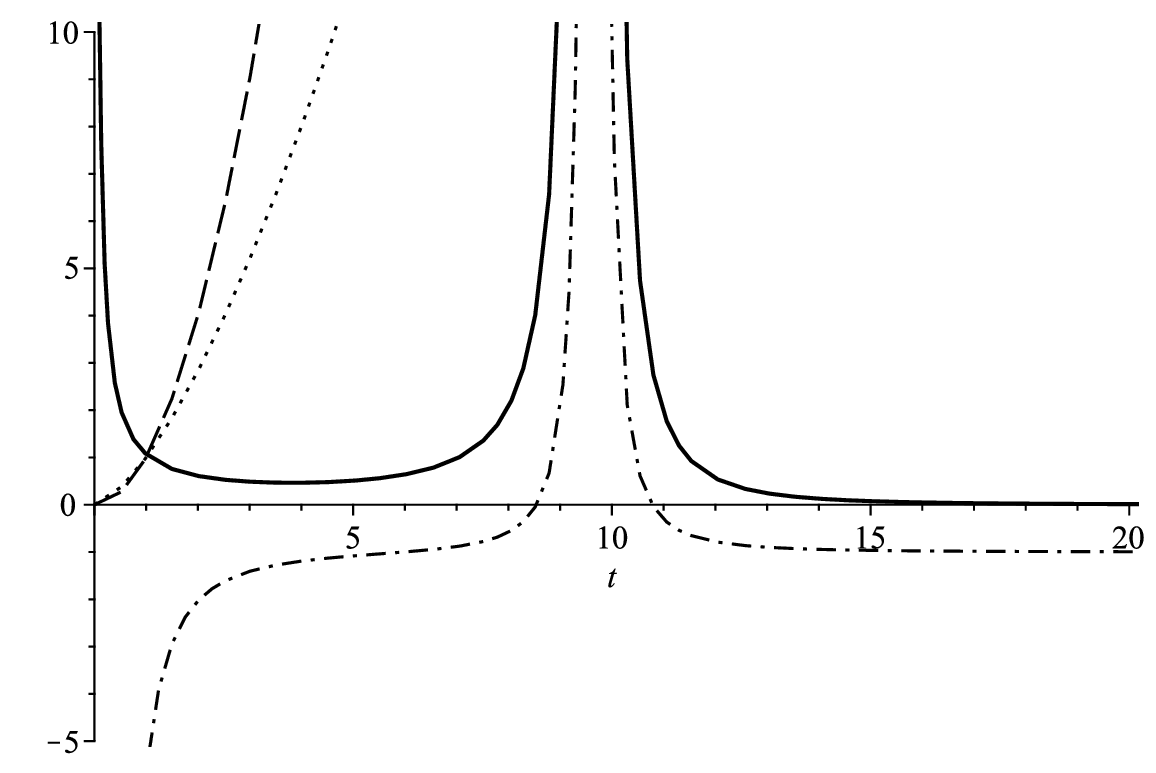}\includegraphics[width=8cm]{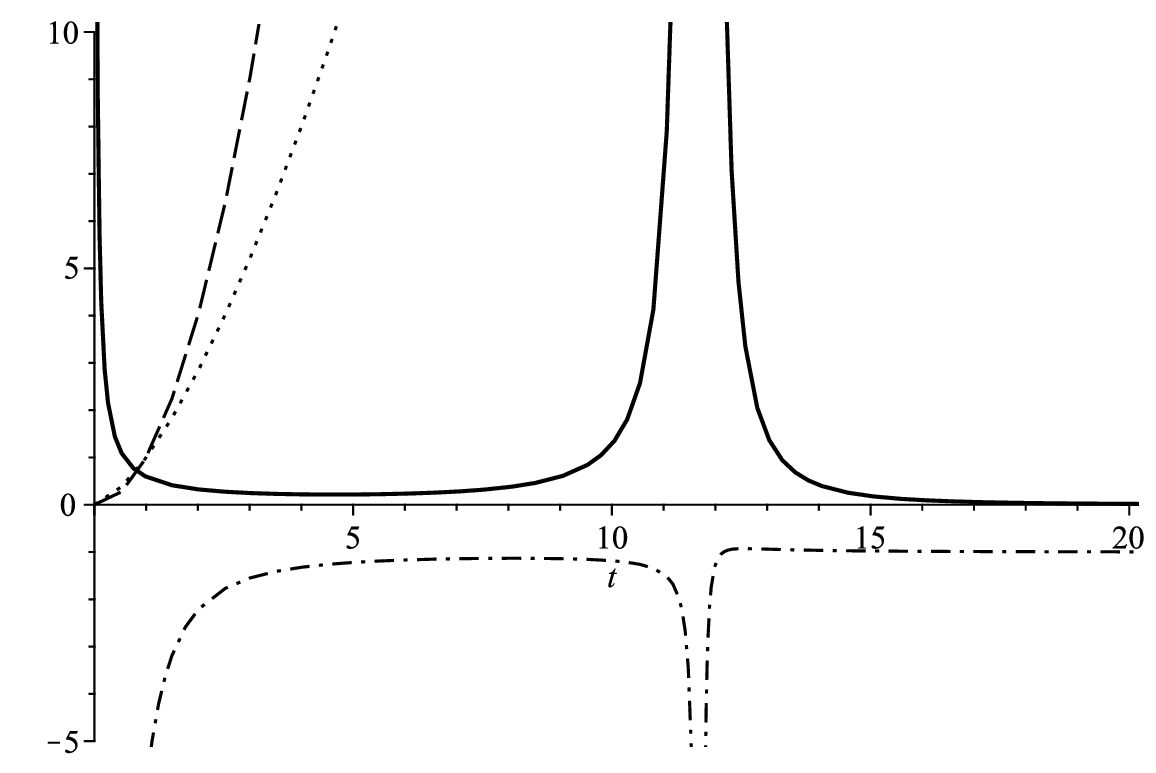}

\caption{For the reconstruction of the power-law cosmology, these figures show
$u\left(t\right)$ (solid line) in Eq.(\ref{eq: u_sol_power}), $\Phi\left(t\right)$
(dashdot line) in Eq.(\ref{eq: Phi_sol_power}), $A\left(t\right)$
(dashed line) and $C\left(t\right)$ (dotted line) in Eq.(\ref{eq: a_c_def_power})
with the parameters $\epsilon=1$, $h_{1}=2$, $h_{2}=1.5$, $C_{1}=3$
and $C_{2}=1$ for left one and the parameters $\epsilon=1$, $h_{1}=0.5$,
$h_{2}=0.4$, $C_{1}=2$ and $C_{2}=0.1$ for the right one.\label{fig: Power_law}}
\end{figure}

\subsection{The Big Rip-like evolution}

The Big Rip is a hypothetical cosmological event that could potentially
end the universe \citep{Caldwell:2003vq,Nojiri:2003vn,Faraoni:2001tq}.
In this scenario, the scale factor, effective energy density, and
effective pressure density diverge. This leads to a universal death,
wherein everything in the universe is progressively torn apart. Let
us consider the following scale factors;

\begin{equation}
A\left(t\right)=\left(t-t_{s}\right)^{a_{0}},\,\,\,\,\,\,\,\,\,\,\,\,\,\,\,C\left(t\right)=\left(t-t_{s}\right)^{c_{0}},\label{eq: a_c_def_bigRip}
\end{equation}
where $a_{0}$ and $c_{0}$ are constants, $t_{0}$ is present age
of the universe. In our cosmological analysis, we have not accounted
for the energy density and pressure terms, focusing instead on the
geometric properties of the underlying space-time. As a result, the
scale factors presented in Eq.(\ref{eq: a_c_def_bigRip}) can be interpreted
as a Big Rip-like evolution which signifies a catastrophic expansion
of the universe.

By the help of Eq.(\ref{eq: a_c_def_bigRip}), the auxiliary metric
looks like 
\begin{equation}
u_{\mu\nu}=u\left(t\right)\text{diag}\left(-1,\left(t-t_{s}\right)^{2a_{0}},\left(t-t_{s}\right)^{2a_{0}},\left(t-t_{s}\right)^{2c_{0}}\right).
\end{equation}

From this background, considering Eq.(\ref{eq: Ricci_metric_relation})
by taking into account of $\left\{ t,t\right\} $, $\left\{ x,x\right\} $,
$\left\{ z,z\right\} $ terms, we find the following three equations,
\begin{equation}
r\left(t\right)=\frac{\left(2a_{0}^{2}+c_{0}^{2}-2a_{0}-c_{0}\right)}{\left(t-t_{s}\right)^{2}}+\frac{H_{u}}{t-t_{s}}\left(a_{0}+\frac{c_{0}}{2}\right)+\frac{3}{2}\dot{H}_{u},\label{eq: r_big_1}
\end{equation}

\begin{equation}
r\left(t\right)=\frac{a_{0}\left(2a_{0}+c_{0}-1\right)}{\left(t-t_{s}\right)^{2}}+\frac{2H_{u}}{t-t_{s}}\left(a_{0}+\frac{c_{0}}{4}\right)+\frac{\ddot{u}}{2u},\label{eq: r_big_2}
\end{equation}

\begin{equation}
r\left(t\right)=\frac{a_{0}\left(2a_{0}+c_{0}-1\right)}{\left(t-t_{s}\right)^{2}}+\frac{H_{u}}{t-t_{s}}\left(a_{0}+\frac{3c_{0}}{2}\right)+\frac{\ddot{u}}{2u},\label{eq: r_big_3}
\end{equation}
and some combination of these equations satisfy

\begin{equation}
\frac{a_{0}\left(c_{0}+1\right)+c_{0}-c_{0}^{2}}{\left(t-t_{s}\right)^{2}}-\frac{a_{0}}{t-t_{s}}H_{u}-\frac{3}{2}H_{u}^{2}+\frac{\ddot{u}}{u}=0,\label{eq: a_0_big}
\end{equation}

\begin{equation}
\frac{2a_{0}\left(a_{0}-c_{0}-1\right)}{\left(t-t_{s}\right)^{2}}-\frac{c_{0}}{t-t_{s}}H_{u}-\frac{3}{2}H_{u}^{2}+\frac{\ddot{u}}{u}=0,\label{eq: c_0_big}
\end{equation}
then solving Eq.(\ref{eq: c_0_big}) with respect to $u\left(t\right)$
we obtain

\begin{equation}
u\left(t\right)=\left[\frac{2\left(t-t_{s}\right)^{-a_{0}}\left(2a_{0}-c_{0}-1\right)}{\left(t-t_{s}\right)^{-2a_{0}+c_{0}+1}-C_{2}}\right]^{2}.\label{eq: u_sol_bigRip}
\end{equation}
Therefore, using Eq.(\ref{eq: r(t)_def}), we can find the curvature
function $f\left(R\right)=\Phi\left(t\right)R$ as

\begin{eqnarray}
\Phi\left(t\right) & = & \frac{1}{\left[C_{1}\left(t-t_{s}\right)^{c_{0}+1}-C_{2}\left(t-t_{s}\right)^{2a_{0}}\right]^{6}}\Biggl\{8\epsilon C_{1}^{5}C_{2}\left(a_{0}-\frac{c_{0}}{4}-\frac{3}{4}\right)\left(t-t_{s}\right)^{2a_{0}+5c_{0}+3}\nonumber \\
 &  & +\epsilon C_{1}^{4}C_{2}^{2}\left[8a_{0}^{2}-33a_{0}c_{0}+17a_{0}+7c_{0}^{2}+19c_{0}-18\right]\left(t-t_{s}\right)^{4a_{0}+4c_{0}+2}\nonumber \\
 &  & -4\epsilon C_{1}^{3}C_{2}^{3}\left[8a_{0}^{2}-13a_{0}c_{0}-3a_{0}+2c_{0}^{2}+9c_{0}-3\right]\left(t-t_{s}\right)^{6a_{0}+3c_{0}+1}\nonumber \\
 &  & +\epsilon C_{1}^{2}C_{2}^{4}\left[28a_{0}^{2}-33a_{0}c_{0}-23a_{0}+2c_{0}^{2}+29c_{0}-3\right]\left(t-t_{s}\right)^{8a_{0}+2c_{0}}\nonumber \\
 &  & +\frac{16\left[C_{1}\left(t-t_{s}\right)^{c_{0}+1}-C_{2}\left(t-t_{s}\right)^{2a_{0}}\right]^{6}}{C_{1}\left(t-t_{s}\right)^{-2a_{0}+c_{0}+1}-C_{2}}\left(a_{0}-\frac{c_{0}}{2}-\frac{1}{2}\right)^{2}\left(t-t_{0}\right)^{-2a_{0}}-\frac{8\epsilon\Upsilon}{\left(t-t_{0}\right)^{2}}\Biggr\}-1,\label{eq: Phi_sol_bigRip}
\end{eqnarray}
where we have used the following expression

\begin{eqnarray}
\Upsilon\left(t\right) & = & \frac{C_{1}^{6}}{2}\left(a_{0}-\frac{c_{0}}{4}-\frac{3}{4}\right)\left(t-t_{s}\right)^{6c_{0}+6}\nonumber \\
 &  & +C_{2}^{5}\left(a_{0}-c_{0}\right)\left[\left(a_{0}+\frac{c_{0}}{4}-\frac{5}{4}\right)\left(t-t_{s}\right)^{10a_{0}+c_{0}+1}-\frac{1}{8}\left(c_{0}-1\right)C_{2}\left(t-t_{0}\right)^{12a_{0}}\right].\label{eq: def_A_big}
\end{eqnarray}

\begin{figure}
\includegraphics[width=8cm]{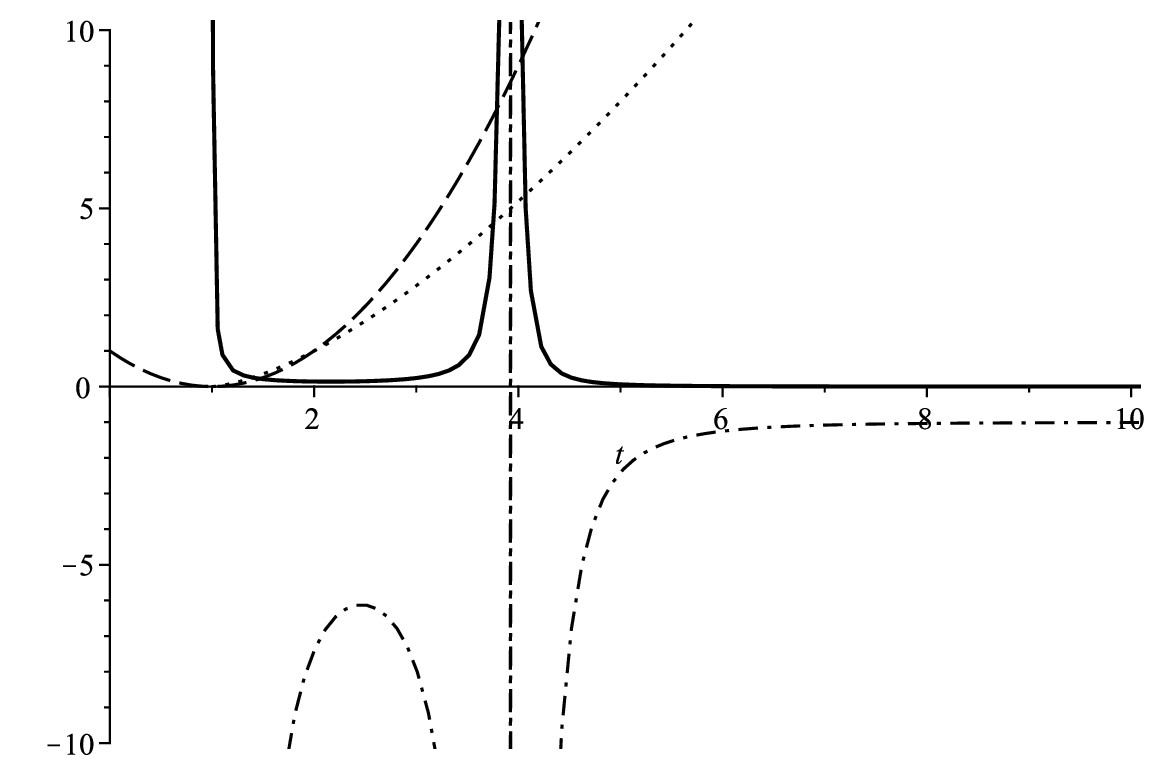}\includegraphics[width=8cm]{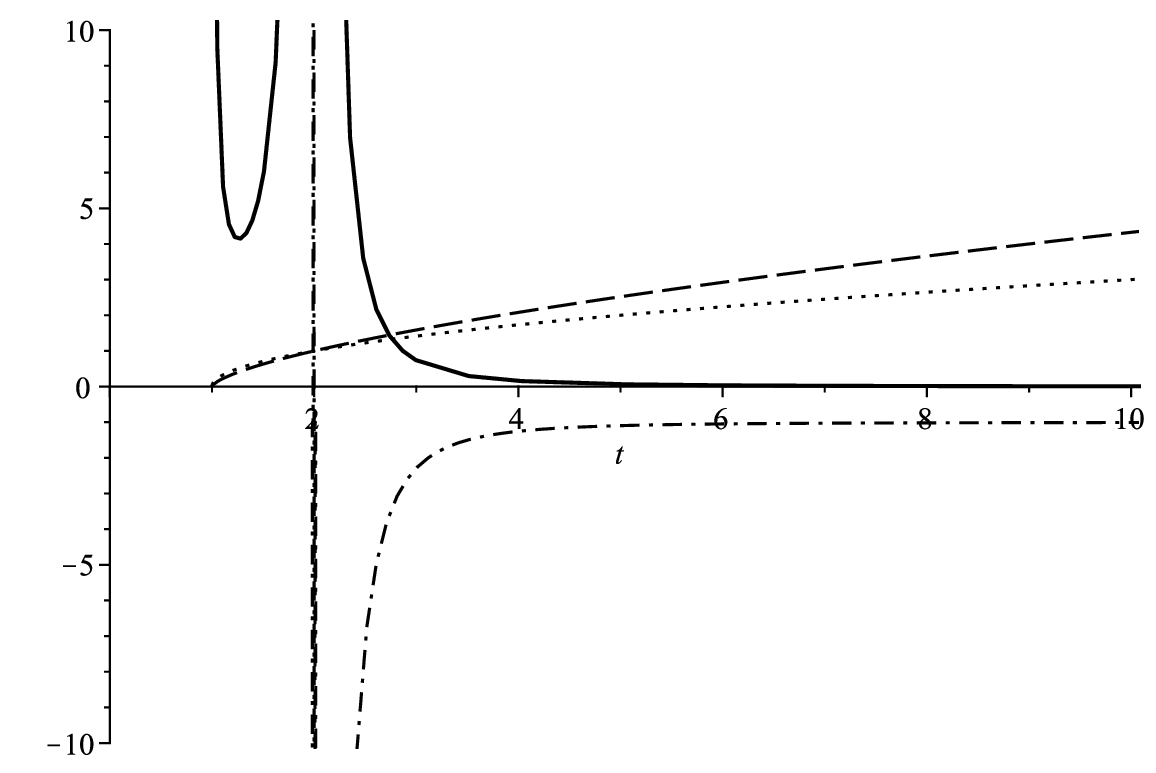}

\caption{With respect to the reconstruction of the Big Rip singularity, these
figures show $u\left(t\right)$ (solid line) in Eq.(\ref{eq: u_sol_bigRip}),
$\Phi\left(t\right)$ (dashdot line) in Eq.(\ref{eq: Phi_sol_bigRip}),
$A\left(t\right)$ (dashed line) and $C\left(t\right)$ (dotted line)
in Eq.(\ref{eq: a_c_def_bigRip}) with the parameters $a_{0}=2$,
$c_{0}=3/2$, $C_{1}=10$ , $C_{2}=2$ and $t_{s}=1$ for left one
and the parameters $a_{0}=2/3$, $c_{0}=1/2$, $C_{1}=2$ , $C_{2}=2$
and $t_{s}=1$ for the right one.\label{fig: big_rip}}
\end{figure}

\subsection{Exponential evolution of $u\left(t\right)$}

Assuming the auxiliary metric function evolves $u\left(t\right)=u_{0}e^{ht}$
where $u_{0}$ and $h$ are chosen to be arbitrary constants thus
the auxiliary metric takes the following form

\begin{equation}
u_{\mu\nu}=u_{0}e^{ht}\text{diag}\left(-1,A\left(t\right)^{2},A\left(t\right)^{2},C\left(t\right)^{2}\right),\label{eq: metric_U-1}
\end{equation}
and using Eq.(\ref{eq: Ricci_metric_relation}) by taking into account
of $\left\{ t,t\right\} $, $\left\{ x,x\right\} $, $\left\{ z,z\right\} $
terms we find the following three equations,
\begin{equation}
r\left(t\right)=\frac{2\ddot{A}}{A}+\frac{\ddot{C}}{C}+hH_{xy}+\frac{h}{2}H_{z},\label{eq: r_eks_u_1}
\end{equation}

\begin{equation}
r\left(t\right)=\frac{h^{2}}{2}+\frac{\ddot{A}}{A}+2hH_{xy}+\frac{h}{2}H_{z}+H_{xy}^{2}+H_{xy}H_{z},\label{eq: r_eks_u_2}
\end{equation}

\begin{equation}
r\left(t\right)=\frac{h^{2}}{2}+\frac{\ddot{C}}{C}+hH_{xy}+\frac{3h}{2}H_{z}+H_{xy}^{2}+2H_{xy}H_{z},\label{eq: r_eks_u_3}
\end{equation}
where $H_{xy}=\frac{\dot{A}}{A}=\frac{\dot{B}}{B}$ is the directional
Hubble parameter along the $x$ and $y$ axis in the axisymmetric
framework. Then some combinations of these equation lead us to write,
\begin{equation}
\frac{\ddot{A}}{A}+\frac{\ddot{C}}{C}-\frac{h^{2}}{2}-hH_{xy}-H_{xy}^{2}-H_{xy}H_{z}=0,\label{eq: a_0}
\end{equation}

\begin{equation}
\frac{2\ddot{A}}{A}-\frac{h^{2}}{2}-hH_{z}-2H_{xy}H_{z}=0.\label{eq: c_0}
\end{equation}
Subsequently, solving Eq.(\ref{eq: c_0}) with respect to $c\left(t\right)$
we find,

\begin{equation}
C\left(t\right)=C_{1}e^{-\frac{ht}{2}}\left(hA\left(t\right)+2\dot{A}\left(t\right)\right),\label{eq: c_sol_1}
\end{equation}
where $C_{1}$ is an integration constant. Then substituting this
solution into Eq.(\ref{eq: a_0}) we obtain the exact expression for
the scale factor $a\left(t\right)$,
\begin{equation}
A\left(t\right)=\frac{1}{2}\left[-\frac{2C_{3}\left(\mathcal{K}^{2}+\mathcal{M}\right)^{2}}{e^{\frac{3ht}{2}}\mathcal{K}^{2}\mathcal{M}}\right]^{\frac{1}{3}},\label{eq: a_sol_exp}
\end{equation}
where we introduce the following expressions for simplifications
\begin{equation}
\mathcal{K}\left(t\right)=e^{\frac{\sqrt{3}e^{\frac{ht}{2}}}{C_{1}h}},\,\,\,\,\,\,\,\,\,\,\,\,\,\,\,\,\,\,\,\,\,\mathcal{M}=e^{\frac{C_{2}\sqrt{3}}{C_{1}}},\label{eq: def_A_and_B}
\end{equation}
here $C_{2}$ and $C_{3}$ are integration constants. Substituting
Eq.(\ref{eq: def_A_and_B}) into Eq.(\ref{eq: c_sol_1}) $c\left(t\right)$
becomes
\begin{equation}
C\left(t\right)=-\frac{2^{1/3}C_{3}\left(\mathcal{K}^{4}-\mathcal{M}^{2}\right)}{\sqrt{3}e^{\frac{3ht}{2}}\mathcal{K}^{2}\mathcal{M}\left\{ -\frac{C_{3}\left(\mathcal{K}^{2}+\mathcal{M}\right)^{2}}{e^{\frac{3ht}{2}}\mathcal{K}^{2}\mathcal{M}}\right\} ^{\frac{2}{3}}}.\label{eq: c_sol_exp}
\end{equation}
From this result one can find the exact expression of the $r\left(t\right)$
function by taking into account of Eqs.(\ref{eq: r_eks_u_1})-(\ref{eq: r_eks_u_3})
\begin{equation}
r\left(t\right)=\frac{e^{ht}}{C_{1}^{2}},\label{eq: r_aks_sol}
\end{equation}
and according to the definition of the auxiliarly metric fuction $u\left(t\right)$
this equation also satisfies $r\left(t\right)=u\left(t\right)/u_{0}C_{1}^{2}$.
Now, equating Eq.(\ref{eq: r(t)_def}) with Eq.(\ref{eq: r_aks_sol})
we get the general definition of $f\left(R\right)$ function as follows
\begin{equation}
f\left(R\right)=\left[\left(u_{0}-\frac{\epsilon}{C_{1}^{2}}\right)e^{ht}-1\right]R+C_{4},
\end{equation}
where $C_{4}$ is the integration constant. 

Having described all the essential features of the Born--Infeld-$f\left(R\right)$
gravity and the corresponding reconstruction method in this particular
case, let us now demonstrate how various cosmological scenarios can
be realized in this model. If one look at the associated Hubble parameters,

\begin{equation}
H_{xy}=-\frac{h}{2}-\frac{e^{\frac{ht}{2}}\left(\mathcal{K}^{2}-\mathcal{M}\right)}{\sqrt{3}C_{1}\left(\mathcal{K}^{2}+\mathcal{M}\right)},\label{eq: H_xy_exp}
\end{equation}

\begin{equation}
H_{z}=-\frac{h}{2}-\frac{e^{ht}\left(\mathcal{K}^{4}+4\mathcal{K}^{2}\mathcal{M}+\mathcal{M}^{2}\right)}{\sqrt{3}C_{1}\left(\mathcal{K}^{4}-\mathcal{M}^{2}\right)}.\label{eq: H_z_exp}
\end{equation}
Therefore the effective equation of state parameter $w_{eff}=-1-\frac{2\dot{H}}{3H^{2}}$
for these scale factors can be found as follows

\begin{equation}
w_{eff}^{xy}=\frac{-9C_{1}^{2}h^{2}\left(\mathcal{K}^{4}+2\mathcal{K}^{2}\mathcal{M}+\mathcal{M}^{2}\right)+\sqrt{3}C_{1}he^{\frac{ht}{2}}\left(\mathcal{K}^{4}-\mathcal{M}^{2}\right)-12e^{ht}\left(\mathcal{K}^{4}+\mathcal{K}^{2}\mathcal{M}+\mathcal{M}^{2}\right)}{\left[3C_{1}h\left(\mathcal{K}^{2}+\mathcal{M}\right)-2\sqrt{3}e^{\frac{ht}{2}}\left(\mathcal{K}^{2}-\mathcal{M}\right)\right]^{2}},\label{eq: w_xy_exp}
\end{equation}

\begin{equation}
w_{eff}^{z}=\frac{8\sqrt{3}C_{1}he^{\frac{ht}{2}}\left[\left(\mathcal{K}^{4}-\mathcal{M}^{2}\right)\left(\mathcal{K}^{4}+4\mathcal{K}^{2}\mathcal{M}+\mathcal{M}^{2}\right)\right]-12e^{ht}\left(\mathcal{K}^{8}+10\mathcal{K}^{4}\mathcal{M}^{2}+\mathcal{M}^{4}\right)-9C_{1}^{2}h^{2}\left(\mathcal{K}^{4}-\mathcal{M}^{2}\right)^{2}}{3\left[\sqrt{3}C_{1}h\left(\mathcal{K}^{4}-\mathcal{M}^{2}\right)-2e^{\frac{ht}{2}}\left(\mathcal{K}^{4}+4\mathcal{K}^{2}\mathcal{M}+\mathcal{M}^{2}\right)\right]^{2}}.\label{eq: w_z_exp}
\end{equation}

\begin{figure}
\includegraphics[width=8cm]{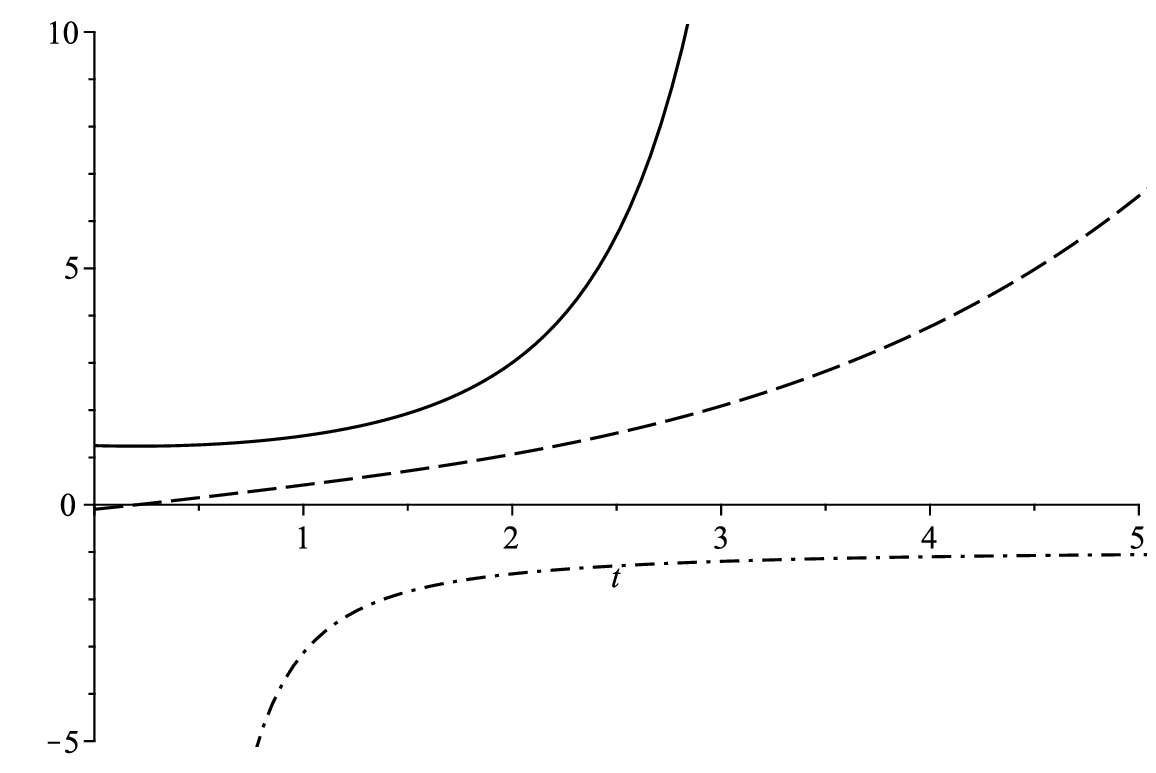}\includegraphics[width=8cm]{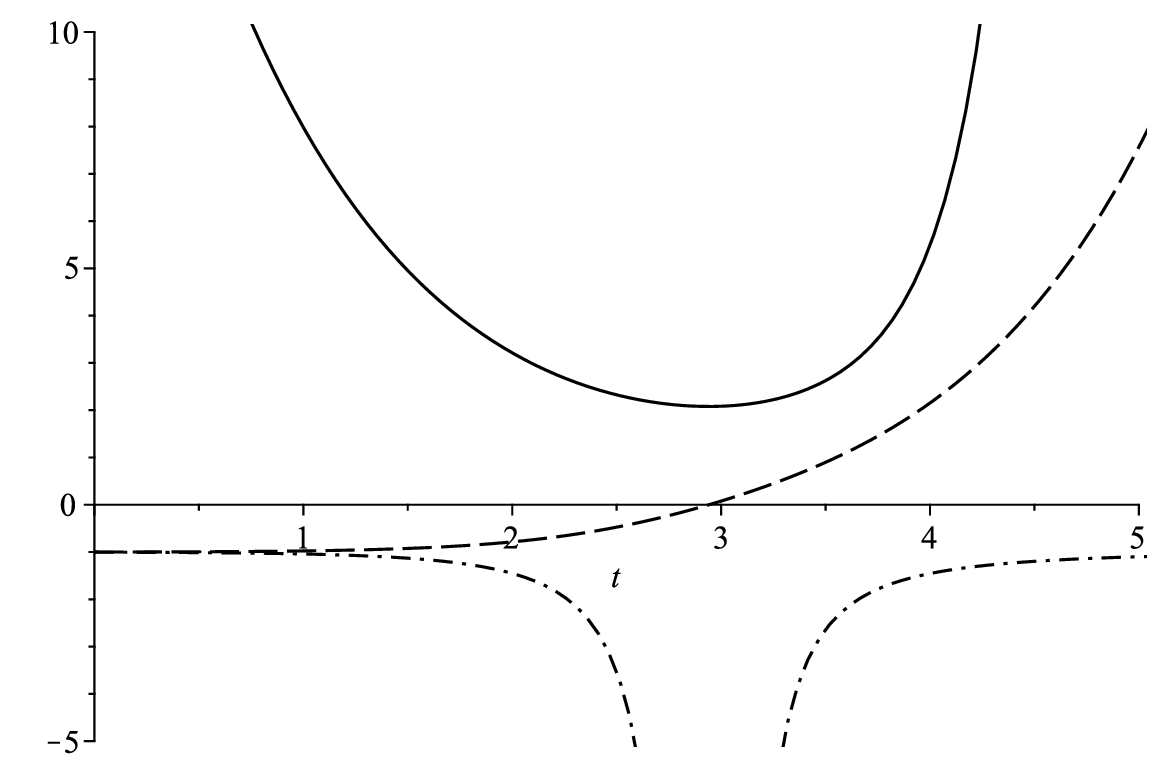}

\caption{Concerning the exponential evolution of the auxiliary metric function
along the $x$ and $y$ axes, these figures show $A\left(t\right)$
(solid line) in Eq.(\ref{eq: a_sol_exp}), $H_{xy}\left(t\right)$
(dashed line) in Eq.(\ref{eq: H_xy_exp}) and $w_{eff}^{xy}\left(t\right)$
(dash-dot line) in Eq.(\ref{eq: w_xy_exp}) with the parameters $h=1$,
$C_{1}=1$, $C_{2}=1$ and $C_{3}=-1$ for left one and the parameters
$h=2$, $C_{1}=10$, $C_{2}=1$ and $C_{3}=-10^{4}$ for the right
one. \label{fig: u_exp_1}}
\end{figure}

\begin{figure}
\includegraphics[width=8cm]{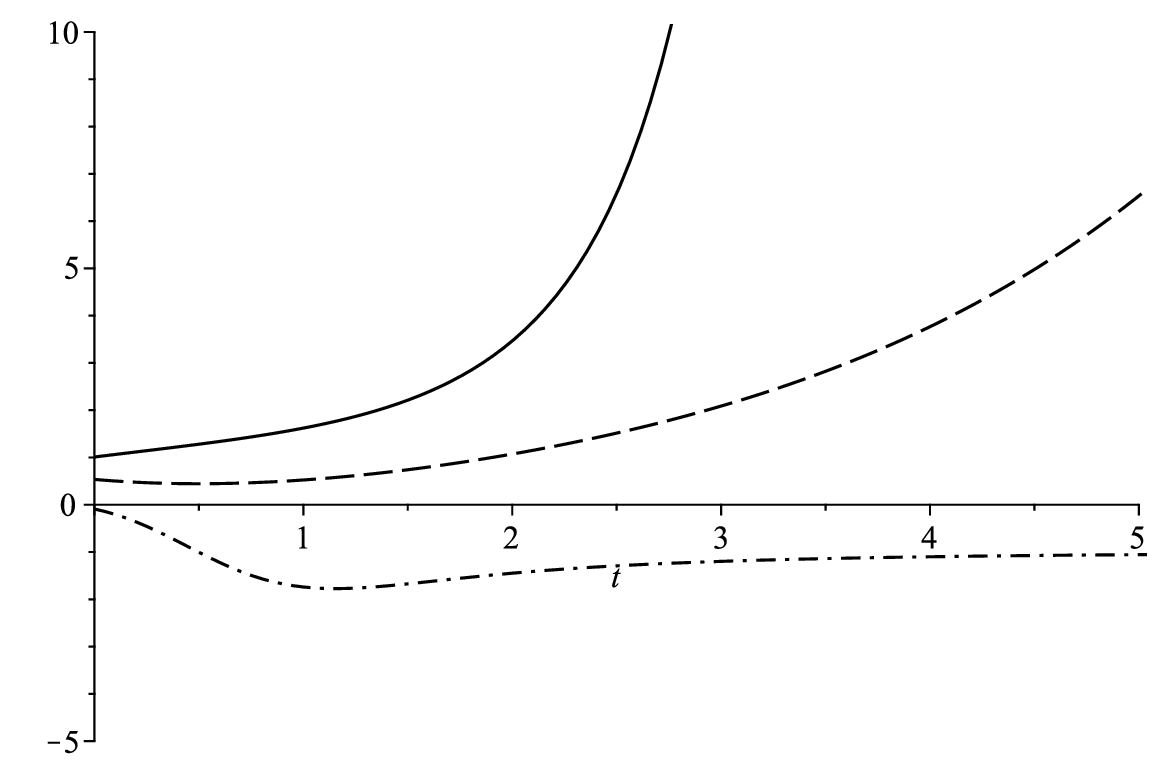}\includegraphics[width=8cm]{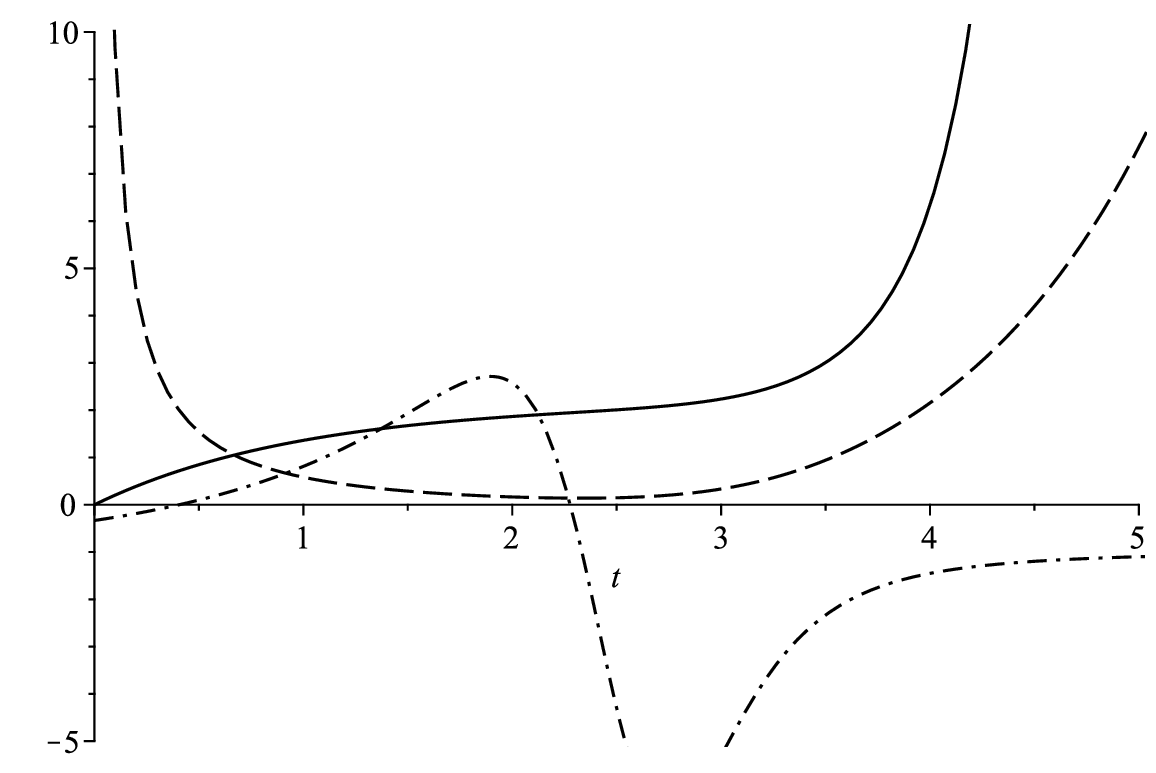}

\caption{In the case of the exponential formulation of the auxiliary metric
function with respect to $z$ axis, these figures show $C\left(t\right)$
(solid line) in Eq.(\ref{eq: c_sol_exp}), $H_{z}\left(t\right)$
(dashed line) in Eq.(\ref{eq: H_z_exp}) and $w_{eff}^{z}\left(t\right)$
(dash-dot line) in Eq.(\ref{eq: w_z_exp}) with the parameters $h=1$,
$C_{1}=1$, $C_{2}=1$ and $C_{3}=-1$ for left one and the parameters
$h=2$, $C_{1}=10$, $C_{2}=1$ and $C_{3}=-10^{4}$ for the right
one. \label{fig: u_exp_2}}
\end{figure}

\subsection{Power-law evolution of of $u\left(t\right)$}

In this case, we suppose that the auxiliary metric function evolves
$u\left(t\right)=t^{h}$ where $h$ is chosen to be arbitrary constants
then the auxiliary metric takes the following form

\begin{equation}
u_{\mu\nu}=t^{h}\text{diag}\left(-1,A\left(t\right)^{2},A\left(t\right)^{2},C\left(t\right)^{2}\right),\label{eq: metric_U-1-1}
\end{equation}
and using Eq.(\ref{eq: Ricci_metric_relation}) by taking into account
of $\left\{ t,t\right\} $, $\left\{ x,x\right\} $, $\left\{ z,z\right\} $
terms we find the following three equations,
\begin{equation}
r\left(t\right)=\frac{2\ddot{A}}{A}+\frac{\ddot{C}}{C}+\frac{h}{t}H_{xy}+\frac{h}{2t}H_{z}-\frac{3h}{2t^{2}},\label{eq: r_aks_1-2}
\end{equation}

\begin{equation}
r\left(t\right)=\frac{\ddot{A}}{A}+\frac{2h}{t}H_{xy}+\frac{h}{2t}H_{z}+H_{xy}^{2}+H_{xy}H_{z}-\frac{h^{2}+h}{2t^{2}},\label{eq: r_aks_2-2}
\end{equation}

\begin{equation}
r\left(t\right)=\frac{\ddot{C}}{C}+\frac{h}{t}H_{xy}+\frac{3h}{2t}H_{z}+2H_{xy}H_{z}-\frac{h^{2}+h}{2t^{2}}.\label{eq: r_aks_3-2}
\end{equation}
Then some combinations of these equation lead us to write,
\begin{equation}
\frac{\ddot{A}}{A}+\frac{\ddot{C}}{C}-\frac{h^{2}+2h}{2t^{2}}-\frac{h}{t}H_{xy}-H_{xy}^{2}-H_{xy}H_{z}=0,\label{eq: sol_1_power}
\end{equation}

\begin{equation}
\frac{2\ddot{A}}{A}-\frac{h^{2}}{2}-\frac{h}{t}H_{z}-2H_{xy}H_{z}-\frac{h^{2}+2h}{2t^{2}}=0.\label{eq: sol_2_power}
\end{equation}
Subsequently, solving Eq.(\ref{eq: sol_2_power}) with respect to
$C\left(t\right)$ we find,

\begin{equation}
C\left(t\right)=C_{1}t^{-\frac{h}{2}}\left(hA\left(t\right)t^{-1}+2\dot{A}\left(t\right)\right),\label{eq: c_func_power}
\end{equation}
where $C_{1}$ is an integration constant. Nevertheless, substituting
Eq.(\ref{eq: c_func_power}) into (\ref{eq: sol_1_power}) it is very
complicated to find the exact expression for the scale factor $A\left(t\right)$.
For this reason, it is helpful to use a specific value for the variable
$h.$ Lets consider a simple example when we set $h=1/2$ then solve
Eq.(\ref{eq: sol_1_power}) taking account of Eq.(\ref{eq: c_func_power})
we find 

\begin{equation}
A\left(t\right)=\frac{e^{C_{3}}}{t^{1/4}}\cosh\left(\mathcal{K}\right),\,\,\,\,\,\,\,\,\,\,\,\,\,\,\,\,\,\,\,\,\,\,\,\,\,\,\,\,C\left(t\right)=\frac{\sqrt{3}e^{C_{3}}}{t^{1/4}}\sinh\left(\mathcal{K}\right),\label{eq: a_c_sol_power}
\end{equation}
where we introduce a time dependent function for simplification:
\begin{equation}
\mathcal{K}\left(t\right)=\frac{\sqrt{3}\left(4t^{5/4}-5C_{2}\right)}{10C_{1}}.
\end{equation}

Therefore the exact expression for the $r\left(t\right)$ function
can be determined by considering Eqs.(\ref{eq: r_eks_u_1})-(\ref{eq: r_eks_u_3})
\begin{equation}
r\left(t\right)=\frac{9\sqrt{t}}{4C_{1}^{2}},\label{eq: r_pow_sol_1}
\end{equation}
Subsequently, we can find the $f\left(R\right)$ expression by the
help of Eqs.(\ref{eq: r(t)_def}) and (\ref{eq: r_pow_sol_1}) as
follows,
\begin{equation}
f\left(R\right)=\left[\frac{\left(4C_{1}^{2}-9\epsilon\right)\sqrt{t}}{4C_{1}^{2}}-1\right]R.
\end{equation}

By examining the associated Hubble parameters
\begin{equation}
H_{xy}=\frac{2\sqrt{3}t^{5/4}\tanh\left(\mathcal{K}\right)-C_{1}}{4C_{1}t},\label{eq: H_xy_power}
\end{equation}

\begin{equation}
H_{z}=\frac{2\sqrt{3}t^{1/4}\coth\left(\mathcal{K}\right)-C_{1}}{4C_{1}t}.\label{eq: H_z_power}
\end{equation}
Thus, the effective equation of state parameter for these scale factors
can be determined as follows
\begin{equation}
w_{eff}^{xy}=-\frac{1}{3}\left(2\coth\left(\mathcal{K}\right)^{2}+\frac{C_{1}t^{5/4}}{\sqrt{3}}\coth\left(\mathcal{K}\right)+1\right),\label{eq: w_xy_power}
\end{equation}

\begin{equation}
w_{eff}^{z}=-\frac{1}{3}\left(2\tanh\left(\mathcal{K}\right)^{2}+\frac{C_{1}t^{5/4}}{\sqrt{3}}\tanh\left(\mathcal{K}\right)+1\right).\label{eq: w_z_power}
\end{equation}

\begin{figure}
\includegraphics[width=8cm]{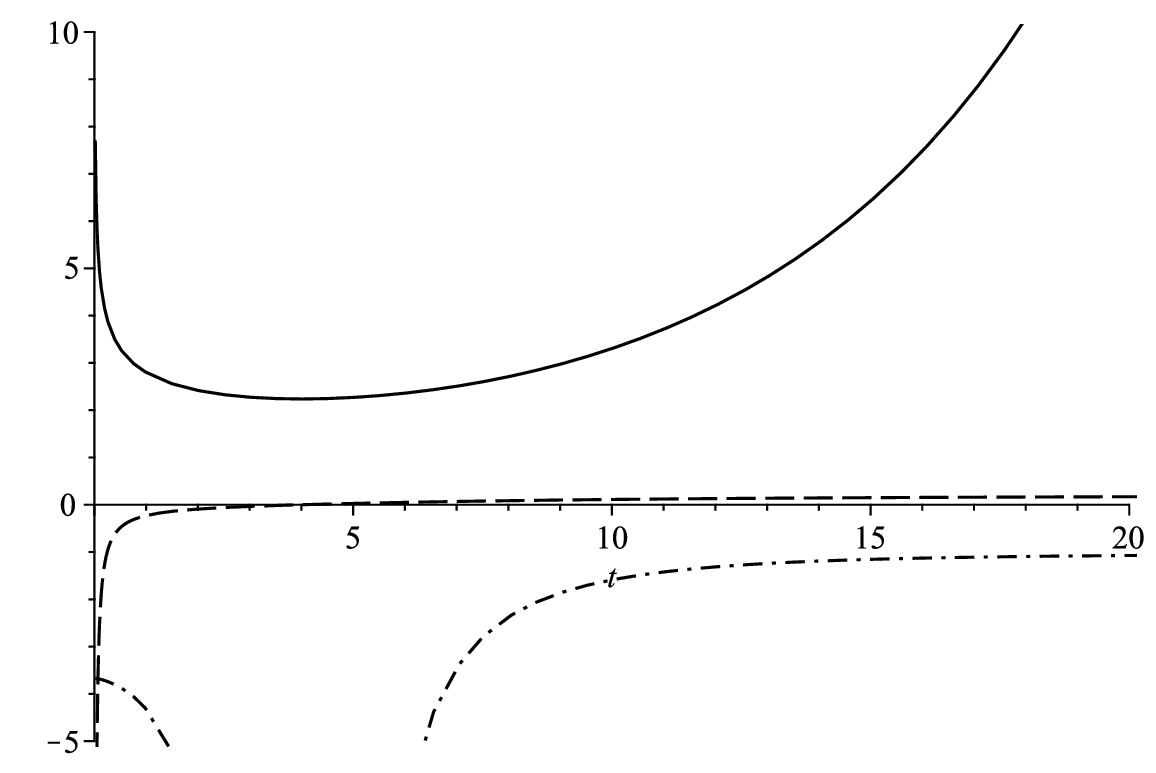}\includegraphics[width=8cm]{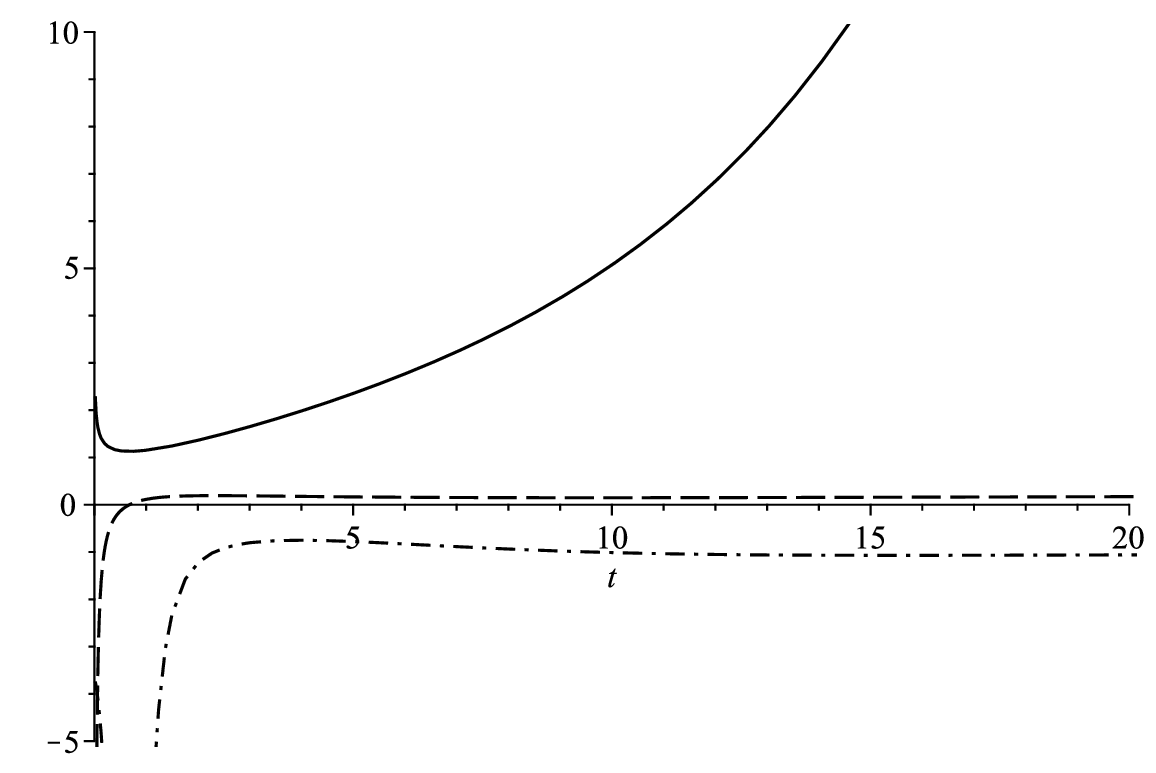}

\caption{Regarding the power-law behavior of the auxiliary metric function,
with the parameters $C_{1}=10$, $C_{2}=-2$ and $C_{3}=1$, the first
figure shows $A\left(t\right)$ (solid line) in Eq.(\ref{eq: a_c_sol_power}),
$H_{xy}\left(t\right)$ (dashed line) in Eq.(\ref{eq: H_xy_power})
and $w_{eff}^{xy}\left(t\right)$ (dash-dot line) in Eq.(\ref{eq: w_xy_power}).
The second one shows $C\left(t\right)$ (solid line) in Eq.(\ref{eq: a_c_sol_power}),
$H_{z}\left(t\right)$ (dashed line) in Eq.(\ref{eq: H_z_power})
and $w_{eff}^{z}\left(t\right)$ (dash-dot line) in Eq.(\ref{eq: w_z_power}).
\label{fig: u_power}}
\end{figure}

\section{\label{sec:Conclusion}Conclusion}

In this study, we have investigated anisotropic cosmological solutions
within the framework of BI-$f\left(R\right)$ gravity, an advanced
modification of general relativity that integrates higher-order curvature
invariants. By focusing on Bianchi type I models, we provided a comprehensive
framework to examine the evolution of spatial anisotropies in the
universe. This approach allows for a detailed exploration of how these
anisotropic properties evolve under the influence of the extended
gravitational theory, offering deeper insights into the fundamental
dynamics of the cosmos.

We analyzed the behavior of Bianchi type I solutions through a reconstruction
scheme, revealing various cosmological scenarios such as de Sitter
like expansion and power-law cosmology. These scenarios offer robust
frameworks for understanding the universe's accelerated expansion,
both during the inflationary period and in the current dark energy-dominated
epoch, within the context of the standard Big Bang evolution. We also
examine the big rip-like singularity, which may provide a cosmological
model ending by the finite time singularity. Examples like these illustrate
the versatility and robustness of the BI-$f\left(R\right)$ gravity
framework in modeling different stages of the universe's evolution
from the early universe to the late-time behavior of the universe
and potential future singularities. 

In addition, we have also analyzed the exponential and power-law evolution
of the auxiliary metric function. In this context, we have found corresponding
cosmological variables such as the scale factors, the Hubble parameters
and the effective equation of state function for corresponding coordinate
axis. Furthermore, by using the reconstruction method we have also
found the corresponding $f\left(R\right)$ functions. According to
these results (look at the figures \ref{fig: u_exp_1}, \ref{fig: u_exp_2},
\ref{fig: u_power}), we can easily say that the formation of the
auxiliary metric function may impact the cosmological dynamics and
lead various dark energy models. As illustrated the first graph in
Fig. \ref{fig: u_exp_1}, this assumption leads to a Little Rip universe
\citep{frampton2011little,frampton2012models,brevik2011viscous,brevik2012turbulence,brevik2012little,makarenko2013big}
($a\left(t\right)\rightarrow\infty$ and $H\left(t\right)\rightarrow\infty$
at future infinity). On the other hand, according to the second graph,
this generalized BI-$f\left(R\right)$ theory can also initiate non-singular
bouncing cosmology \citep{Peter:2002Primordial,Allen:2004Cosmological,Cai:2011MatterBounce,Cai:2012Towards,Nojiri:2019NonsingularBounce,Odintsov:2020FromBounce}
when the auxiliary metric function takes certain forms. These forms
are unconventional within standard Einstein-Hilbert gravity and cannot
be achieved in that framework. The remaining cases explicitly demonstrates
that the effective EoS approaches to $-1$ for the late time era so
this behavior is similar to the cosmological constant ($w_{\Lambda}=-1$).

Our analysis demonstrates that the modifications introduced by BI-$f\left(R\right)$
gravity significantly impact anisotropic dynamics, offering new perspectives
on the late-time behavior of the universe and possible future singularities
\citep{nojiri2005properties,deHaro:2023Finite-time,Katsuragawa:2024FutureSingularity}.
We note that it would be intriguing to explore the unification of
various cosmological eras within BI-$f\left(R\right)$ gravity framework
in the anisotropic condition, which we plan to investigate in future
work. We also note that BI-$f\left(R\right)$ gravity hasn't been
extensively explored in the context of the Hubble tension, $H_{0}$
\citep{Bernal:2016TheTroubleH0}. In this context, the recent review
\citep{Vagnozzi:2023Seven} is devoted to the study of modifications
of early and late universe descriptions to solve the current cosmic
tensions. This will be done elsewhere. Moreover, the theory under
consideration exhibits characteristics analogous to a conformal transformation,
as expressed in Eq.(\ref{eq: conf_rel}), which defines a conformal
relationship between the metrics $g_{\mu\nu}$ and $q_{\mu\nu}$ but
this relation does not correspond to the well-known conformal transformation
for the metric tensor \citep{jimenez2018born}. The relation between
the scalar-tensor and $f\left(R\right)$ theories motivates us to
say that it is interesting to analyze the conformal transformation
of the space-time metric tensor to determine the resulting theoretical
framework. This analysis is expected to reveal increased complexity
following the conformal transformation, necessitating a thorough examination
of the emergent dynamics and interactions within the transformed theory.
We intend to address some of the aforementioned issues in a subsequent
work.

\section*{Declaration of competing interest }

The authors declare that they have no known competing financial interests
or personal relationships that could have appeared to influence the
work reported in this paper.

\section*{Declaration of generative AI and AI-assisted technologies in the
writing process}

During the preparation of this work the author used ChatGPT - GPT-4o
in order to improve readability of the text. After using this tool/service,
the author reviewed and edited the content as needed and takes full
responsibility for the content of the publication.

\section*{Data availability }

No data was used for the research described in the article.

\bibliography{BIFR_Anisotropic}

\begin{thebibliography}{89}%
\makeatletter
\providecommand \@ifxundefined [1]{%
 \@ifx{#1\undefined}
}%
\providecommand \@ifnum [1]{%
 \ifnum #1\expandafter \@firstoftwo
 \else \expandafter \@secondoftwo
 \fi
}%
\providecommand \@ifx [1]{%
 \ifx #1\expandafter \@firstoftwo
 \else \expandafter \@secondoftwo
 \fi
}%
\providecommand \natexlab [1]{#1}%
\providecommand \enquote  [1]{``#1''}%
\providecommand \bibnamefont  [1]{#1}%
\providecommand \bibfnamefont [1]{#1}%
\providecommand \citenamefont [1]{#1}%
\providecommand \href@noop [0]{\@secondoftwo}%
\providecommand \href [0]{\begingroup \@sanitize@url \@href}%
\providecommand \@href[1]{\@@startlink{#1}\@@href}%
\providecommand \@@href[1]{\endgroup#1\@@endlink}%
\providecommand \@sanitize@url [0]{\catcode `\\12\catcode `\$12\catcode
  `\&12\catcode `\#12\catcode `\^12\catcode `\_12\catcode `\%12\relax}%
\providecommand \@@startlink[1]{}%
\providecommand \@@endlink[0]{}%
\providecommand \url  [0]{\begingroup\@sanitize@url \@url }%
\providecommand \@url [1]{\endgroup\@href {#1}{\urlprefix }}%
\providecommand \urlprefix  [0]{URL }%
\providecommand \Eprint [0]{\href }%
\providecommand \doibase [0]{https://doi.org/}%
\providecommand \selectlanguage [0]{\@gobble}%
\providecommand \bibinfo  [0]{\@secondoftwo}%
\providecommand \bibfield  [0]{\@secondoftwo}%
\providecommand \translation [1]{[#1]}%
\providecommand \BibitemOpen [0]{}%
\providecommand \bibitemStop [0]{}%
\providecommand \bibitemNoStop [0]{.\EOS\space}%
\providecommand \EOS [0]{\spacefactor3000\relax}%
\providecommand \BibitemShut  [1]{\csname bibitem#1\endcsname}%
\let\auto@bib@innerbib\@empty
\bibitem [{\citenamefont {Nojiri}\ and\ \citenamefont
  {Odintsov}(2011)}]{nojiri2011unified}%
  \BibitemOpen
  \bibfield  {author} {\bibinfo {author} {\bibfnamefont {S.}~\bibnamefont
  {Nojiri}}\ and\ \bibinfo {author} {\bibfnamefont {S.~D.}\ \bibnamefont
  {Odintsov}},\ }\bibfield  {title} {\bibinfo {title} {{Unified cosmic history
  in modified gravity: from F(R) theory to Lorentz non-invariant models}},\
  }\href {https://doi.org/10.1016/j.physrep.2011.04.001} {\bibfield  {journal}
  {\bibinfo  {journal} {Phys. Rept.}\ }\textbf {\bibinfo {volume} {505}},\
  \bibinfo {pages} {59} (\bibinfo {year} {2011})}\BibitemShut {NoStop}%
\bibitem [{\citenamefont {Nojiri}\ and\ \citenamefont
  {Odintsov}(2007)}]{nojiri2007introduction}%
  \BibitemOpen
  \bibfield  {author} {\bibinfo {author} {\bibfnamefont {S.}~\bibnamefont
  {Nojiri}}\ and\ \bibinfo {author} {\bibfnamefont {S.~D.}\ \bibnamefont
  {Odintsov}},\ }\bibfield  {title} {\bibinfo {title} {Introduction to modified
  gravity and gravitational alternative for dark energy},\ }\href
  {https://doi.org/10.1142/S0219887807001928} {\bibfield  {journal} {\bibinfo
  {journal} {Int. J. Geom. Methods Mod. Phys.}\ }\textbf {\bibinfo {volume}
  {4}},\ \bibinfo {pages} {115} (\bibinfo {year} {2007})}\BibitemShut {NoStop}%
\bibitem [{\citenamefont {Nojiri}\ \emph {et~al.}(2017)\citenamefont {Nojiri},
  \citenamefont {Odintsov},\ and\ \citenamefont {Oikonomou}}]{Nojiri:2017ncd}%
  \BibitemOpen
  \bibfield  {author} {\bibinfo {author} {\bibfnamefont {S.}~\bibnamefont
  {Nojiri}}, \bibinfo {author} {\bibfnamefont {S.~D.}\ \bibnamefont
  {Odintsov}},\ and\ \bibinfo {author} {\bibfnamefont {V.~K.}\ \bibnamefont
  {Oikonomou}},\ }\bibfield  {title} {\bibinfo {title} {{Modified Gravity
  Theories on a Nutshell: Inflation, Bounce and Late-time Evolution}},\ }\href
  {https://doi.org/10.1016/j.physrep.2017.06.001} {\bibfield  {journal}
  {\bibinfo  {journal} {Phys. Rept.}\ }\textbf {\bibinfo {volume} {692}},\
  \bibinfo {pages} {1} (\bibinfo {year} {2017})}\BibitemShut {NoStop}%
\bibitem [{\citenamefont {Born}\ and\ \citenamefont
  {Infeld}(1934)}]{born1934foundations}%
  \BibitemOpen
  \bibfield  {author} {\bibinfo {author} {\bibfnamefont {M.}~\bibnamefont
  {Born}}\ and\ \bibinfo {author} {\bibfnamefont {L.}~\bibnamefont {Infeld}},\
  }\bibfield  {title} {\bibinfo {title} {{Foundations of the new field
  theory}},\ }\href {https://doi.org/10.1098/rspa.1934.0059} {\bibfield
  {journal} {\bibinfo  {journal} {Proc. Roy. Soc. Lond. A}\ }\textbf {\bibinfo
  {volume} {144}},\ \bibinfo {pages} {425} (\bibinfo {year}
  {1934})}\BibitemShut {NoStop}%
\bibitem [{\citenamefont {Deser}\ and\ \citenamefont
  {Gibbons}(1998)}]{deser1998born}%
  \BibitemOpen
  \bibfield  {author} {\bibinfo {author} {\bibfnamefont {S.}~\bibnamefont
  {Deser}}\ and\ \bibinfo {author} {\bibfnamefont {G.~W.}\ \bibnamefont
  {Gibbons}},\ }\bibfield  {title} {\bibinfo {title} {{Born-Infeld-Einstein
  actions?}},\ }\href {https://doi.org/10.1088/0264-9381/15/5/001} {\bibfield
  {journal} {\bibinfo  {journal} {Class. Quant. Grav.}\ }\textbf {\bibinfo
  {volume} {15}},\ \bibinfo {pages} {L35} (\bibinfo {year} {1998})}\BibitemShut
  {NoStop}%
\bibitem [{\citenamefont {Eddington}(1923)}]{eddington1923mathematical}%
  \BibitemOpen
  \bibfield  {author} {\bibinfo {author} {\bibfnamefont {A.~S.}\ \bibnamefont
  {Eddington}},\ }\href@noop {} {\emph {\bibinfo {title} {The Mathematical
  Theory of Relativity}}}\ (\bibinfo  {publisher} {The Cambridge University
  Press},\ \bibinfo {year} {1923})\BibitemShut {NoStop}%
\bibitem [{\citenamefont {Vollick}(2004)}]{vollick2004palatini}%
  \BibitemOpen
  \bibfield  {author} {\bibinfo {author} {\bibfnamefont {D.~N.}\ \bibnamefont
  {Vollick}},\ }\bibfield  {title} {\bibinfo {title} {{Palatini approach to
  Born-Infeld-Einstein theory and a geometric description of
  electrodynamics}},\ }\href {https://doi.org/10.1103/PhysRevD.69.064030}
  {\bibfield  {journal} {\bibinfo  {journal} {Phys. Rev. D}\ }\textbf {\bibinfo
  {volume} {69}},\ \bibinfo {pages} {064030} (\bibinfo {year}
  {2004})}\BibitemShut {NoStop}%
\bibitem [{\citenamefont {Vollick}(2005)}]{vollick2005born}%
  \BibitemOpen
  \bibfield  {author} {\bibinfo {author} {\bibfnamefont {D.~N.}\ \bibnamefont
  {Vollick}},\ }\bibfield  {title} {\bibinfo {title} {{Born-Infeld-Einstein
  theory with matter}},\ }\href {https://doi.org/10.1103/PhysRevD.72.084026}
  {\bibfield  {journal} {\bibinfo  {journal} {Phys. Rev. D}\ }\textbf {\bibinfo
  {volume} {72}},\ \bibinfo {pages} {084026} (\bibinfo {year}
  {2005})}\BibitemShut {NoStop}%
\bibitem [{\citenamefont {Ba\~nados}\ and\ \citenamefont
  {Ferreira}(2010)}]{Banados2010eddington}%
  \BibitemOpen
  \bibfield  {author} {\bibinfo {author} {\bibfnamefont {M.}~\bibnamefont
  {Ba\~nados}}\ and\ \bibinfo {author} {\bibfnamefont {P.~G.}\ \bibnamefont
  {Ferreira}},\ }\bibfield  {title} {\bibinfo {title} {Eddington's theory of
  gravity and its progeny},\ }\href
  {https://doi.org/10.1103/PhysRevLett.105.011101} {\bibfield  {journal}
  {\bibinfo  {journal} {Phys. Rev. Lett.}\ }\textbf {\bibinfo {volume} {105}},\
  \bibinfo {pages} {011101} (\bibinfo {year} {2010})},\ \bibinfo {note}
  {[Erratum: Phys.Rev.Lett. 113, 119901 (2014)]}\BibitemShut {NoStop}%
\bibitem [{\citenamefont {Avelino}\ and\ \citenamefont
  {Ferreira}(2012)}]{Avelino:2012BIcosmologies}%
  \BibitemOpen
  \bibfield  {author} {\bibinfo {author} {\bibfnamefont {P.~P.}\ \bibnamefont
  {Avelino}}\ and\ \bibinfo {author} {\bibfnamefont {R.~Z.}\ \bibnamefont
  {Ferreira}},\ }\bibfield  {title} {\bibinfo {title} {{Bouncing
  Eddington-inspired Born-Infeld cosmologies: an alternative to Inflation ?}},\
  }\href {https://doi.org/10.1103/PhysRevD.86.041501} {\bibfield  {journal}
  {\bibinfo  {journal} {Phys. Rev. D}\ }\textbf {\bibinfo {volume} {86}},\
  \bibinfo {pages} {041501} (\bibinfo {year} {2012})}\BibitemShut {NoStop}%
\bibitem [{\citenamefont {Escamilla-Rivera}\ \emph {et~al.}(2012)\citenamefont
  {Escamilla-Rivera}, \citenamefont {Banados},\ and\ \citenamefont
  {Ferreira}}]{Escamilla-Rivera:2012Tensor}%
  \BibitemOpen
  \bibfield  {author} {\bibinfo {author} {\bibfnamefont {C.}~\bibnamefont
  {Escamilla-Rivera}}, \bibinfo {author} {\bibfnamefont {M.}~\bibnamefont
  {Banados}},\ and\ \bibinfo {author} {\bibfnamefont {P.~G.}\ \bibnamefont
  {Ferreira}},\ }\bibfield  {title} {\bibinfo {title} {{A tensor instability in
  the Eddington inspired Born-Infeld Theory of Gravity}},\ }\href
  {https://doi.org/10.1103/PhysRevD.85.087302} {\bibfield  {journal} {\bibinfo
  {journal} {Phys. Rev. D}\ }\textbf {\bibinfo {volume} {85}},\ \bibinfo
  {pages} {087302} (\bibinfo {year} {2012})}\BibitemShut {NoStop}%
\bibitem [{\citenamefont {Cho}\ \emph {et~al.}(2012)\citenamefont {Cho},
  \citenamefont {Kim},\ and\ \citenamefont {Moon}}]{Cho:2012Universe}%
  \BibitemOpen
  \bibfield  {author} {\bibinfo {author} {\bibfnamefont {I.}~\bibnamefont
  {Cho}}, \bibinfo {author} {\bibfnamefont {H.-C.}\ \bibnamefont {Kim}},\ and\
  \bibinfo {author} {\bibfnamefont {T.}~\bibnamefont {Moon}},\ }\bibfield
  {title} {\bibinfo {title} {{Universe Driven by Perfect Fluid in
  Eddington-inspired Born-Infeld Gravity}},\ }\href
  {https://doi.org/10.1103/PhysRevD.86.084018} {\bibfield  {journal} {\bibinfo
  {journal} {Phys. Rev. D}\ }\textbf {\bibinfo {volume} {86}},\ \bibinfo
  {pages} {084018} (\bibinfo {year} {2012})}\BibitemShut {NoStop}%
\bibitem [{\citenamefont {Scargill}\ \emph {et~al.}(2012)\citenamefont
  {Scargill}, \citenamefont {Banados},\ and\ \citenamefont
  {Ferreira}}]{Scargill:2012Cosmology}%
  \BibitemOpen
  \bibfield  {author} {\bibinfo {author} {\bibfnamefont {J.~H.~C.}\
  \bibnamefont {Scargill}}, \bibinfo {author} {\bibfnamefont {M.}~\bibnamefont
  {Banados}},\ and\ \bibinfo {author} {\bibfnamefont {P.~G.}\ \bibnamefont
  {Ferreira}},\ }\bibfield  {title} {\bibinfo {title} {{Cosmology with
  Eddington-inspired Gravity}},\ }\href
  {https://doi.org/10.1103/PhysRevD.86.103533} {\bibfield  {journal} {\bibinfo
  {journal} {Phys. Rev. D}\ }\textbf {\bibinfo {volume} {86}},\ \bibinfo
  {pages} {103533} (\bibinfo {year} {2012})}\BibitemShut {NoStop}%
\bibitem [{\citenamefont {Kruglov}(2014)}]{Kruglov:2013Modified}%
  \BibitemOpen
  \bibfield  {author} {\bibinfo {author} {\bibfnamefont {S.~I.}\ \bibnamefont
  {Kruglov}},\ }\bibfield  {title} {\bibinfo {title} {{Modified arctan-gravity
  model mimicking a cosmological constant}},\ }\href
  {https://doi.org/10.1103/PhysRevD.89.064004} {\bibfield  {journal} {\bibinfo
  {journal} {Phys. Rev. D}\ }\textbf {\bibinfo {volume} {89}},\ \bibinfo
  {pages} {064004} (\bibinfo {year} {2014})}\BibitemShut {NoStop}%
\bibitem [{\citenamefont {Yang}\ \emph {et~al.}(2013)\citenamefont {Yang},
  \citenamefont {Du},\ and\ \citenamefont {Liu}}]{Yang:2013Linear}%
  \BibitemOpen
  \bibfield  {author} {\bibinfo {author} {\bibfnamefont {K.}~\bibnamefont
  {Yang}}, \bibinfo {author} {\bibfnamefont {X.-L.}\ \bibnamefont {Du}},\ and\
  \bibinfo {author} {\bibfnamefont {Y.-X.}\ \bibnamefont {Liu}},\ }\bibfield
  {title} {\bibinfo {title} {{Linear perturbations in Eddington-inspired
  Born-Infeld gravity}},\ }\href {https://doi.org/10.1103/PhysRevD.88.124037}
  {\bibfield  {journal} {\bibinfo  {journal} {Phys. Rev. D}\ }\textbf {\bibinfo
  {volume} {88}},\ \bibinfo {pages} {124037} (\bibinfo {year}
  {2013})}\BibitemShut {NoStop}%
\bibitem [{\citenamefont {Du}\ \emph {et~al.}(2014)\citenamefont {Du},
  \citenamefont {Yang}, \citenamefont {Meng},\ and\ \citenamefont
  {Liu}}]{Du:2014Large}%
  \BibitemOpen
  \bibfield  {author} {\bibinfo {author} {\bibfnamefont {X.-L.}\ \bibnamefont
  {Du}}, \bibinfo {author} {\bibfnamefont {K.}~\bibnamefont {Yang}}, \bibinfo
  {author} {\bibfnamefont {X.-H.}\ \bibnamefont {Meng}},\ and\ \bibinfo
  {author} {\bibfnamefont {Y.-X.}\ \bibnamefont {Liu}},\ }\bibfield  {title}
  {\bibinfo {title} {{Large Scale Structure Formation in Eddington-inspired
  Born-Infeld Gravity}},\ }\href {https://doi.org/10.1103/PhysRevD.90.044054}
  {\bibfield  {journal} {\bibinfo  {journal} {Phys. Rev. D}\ }\textbf {\bibinfo
  {volume} {90}},\ \bibinfo {pages} {044054} (\bibinfo {year}
  {2014})}\BibitemShut {NoStop}%
\bibitem [{\citenamefont {Kim}(2014)}]{Kim:2014Origin}%
  \BibitemOpen
  \bibfield  {author} {\bibinfo {author} {\bibfnamefont {H.-C.}\ \bibnamefont
  {Kim}},\ }\bibfield  {title} {\bibinfo {title} {{Origin of the universe: A
  hint from Eddington-inspired Born-Infeld gravity}},\ }\href
  {https://doi.org/10.3938/jkps.65.840} {\bibfield  {journal} {\bibinfo
  {journal} {J. Korean Phys. Soc.}\ }\textbf {\bibinfo {volume} {65}},\
  \bibinfo {pages} {840} (\bibinfo {year} {2014})}\BibitemShut {NoStop}%
\bibitem [{\citenamefont {Harko}\ \emph {et~al.}(2015)\citenamefont {Harko},
  \citenamefont {Lobo}, \citenamefont {Mak},\ and\ \citenamefont
  {Sushkov}}]{Harko:2015Wormhole}%
  \BibitemOpen
  \bibfield  {author} {\bibinfo {author} {\bibfnamefont {T.}~\bibnamefont
  {Harko}}, \bibinfo {author} {\bibfnamefont {F.~S.~N.}\ \bibnamefont {Lobo}},
  \bibinfo {author} {\bibfnamefont {M.~K.}\ \bibnamefont {Mak}},\ and\ \bibinfo
  {author} {\bibfnamefont {S.~V.}\ \bibnamefont {Sushkov}},\ }\bibfield
  {title} {\bibinfo {title} {{Wormhole geometries in Eddington-Inspired
  Born\textendash{}Infeld gravity}},\ }\href
  {https://doi.org/10.1142/S0217732315501904} {\bibfield  {journal} {\bibinfo
  {journal} {Mod. Phys. Lett. A}\ }\textbf {\bibinfo {volume} {30}},\ \bibinfo
  {pages} {1550190} (\bibinfo {year} {2015})}\BibitemShut {NoStop}%
\bibitem [{\citenamefont {Olmo}\ \emph {et~al.}(2014)\citenamefont {Olmo},
  \citenamefont {Rubiera-Garcia},\ and\ \citenamefont
  {Sanchis-Alepuz}}]{Olmo:2014GeonicBH}%
  \BibitemOpen
  \bibfield  {author} {\bibinfo {author} {\bibfnamefont {G.~J.}\ \bibnamefont
  {Olmo}}, \bibinfo {author} {\bibfnamefont {D.}~\bibnamefont
  {Rubiera-Garcia}},\ and\ \bibinfo {author} {\bibfnamefont {H.}~\bibnamefont
  {Sanchis-Alepuz}},\ }\bibfield  {title} {\bibinfo {title} {{Geonic black
  holes and remnants in Eddington-inspired Born-Infeld gravity}},\ }\href
  {https://doi.org/10.1140/epjc/s10052-014-2804-8} {\bibfield  {journal}
  {\bibinfo  {journal} {Eur. Phys. J. C}\ }\textbf {\bibinfo {volume} {74}},\
  \bibinfo {pages} {2804} (\bibinfo {year} {2014})}\BibitemShut {NoStop}%
\bibitem [{\citenamefont {Lobo}\ \emph {et~al.}(2014)\citenamefont {Lobo},
  \citenamefont {Olmo},\ and\ \citenamefont
  {Rubiera-Garcia}}]{Lobo:2014Microscopic}%
  \BibitemOpen
  \bibfield  {author} {\bibinfo {author} {\bibfnamefont {F.~S.~N.}\
  \bibnamefont {Lobo}}, \bibinfo {author} {\bibfnamefont {G.~J.}\ \bibnamefont
  {Olmo}},\ and\ \bibinfo {author} {\bibfnamefont {D.}~\bibnamefont
  {Rubiera-Garcia}},\ }\bibfield  {title} {\bibinfo {title} {{Microscopic
  wormholes and the geometry of entanglement}},\ }\href
  {https://doi.org/10.1140/epjc/s10052-014-2924-1} {\bibfield  {journal}
  {\bibinfo  {journal} {Eur. Phys. J. C}\ }\textbf {\bibinfo {volume} {74}},\
  \bibinfo {pages} {2924} (\bibinfo {year} {2014})}\BibitemShut {NoStop}%
\bibitem [{\citenamefont {Makarenko}\ \emph
  {et~al.}(2014{\natexlab{a}})\citenamefont {Makarenko}, \citenamefont
  {Odintsov},\ and\ \citenamefont {Olmo}}]{makarenko2014born}%
  \BibitemOpen
  \bibfield  {author} {\bibinfo {author} {\bibfnamefont {A.~N.}\ \bibnamefont
  {Makarenko}}, \bibinfo {author} {\bibfnamefont {S.~D.}\ \bibnamefont
  {Odintsov}},\ and\ \bibinfo {author} {\bibfnamefont {G.~J.}\ \bibnamefont
  {Olmo}},\ }\bibfield  {title} {\bibinfo {title} {{Born-Infeld-$f(R)$
  gravity}},\ }\href {https://doi.org/10.1103/PhysRevD.90.024066} {\bibfield
  {journal} {\bibinfo  {journal} {Phys. Rev. D}\ }\textbf {\bibinfo {volume}
  {90}},\ \bibinfo {pages} {024066} (\bibinfo {year}
  {2014}{\natexlab{a}})}\BibitemShut {NoStop}%
\bibitem [{\citenamefont {Makarenko}\ \emph
  {et~al.}(2014{\natexlab{b}})\citenamefont {Makarenko}, \citenamefont
  {Odintsov},\ and\ \citenamefont {Olmo}}]{Makarenko:2014nca}%
  \BibitemOpen
  \bibfield  {author} {\bibinfo {author} {\bibfnamefont {A.~N.}\ \bibnamefont
  {Makarenko}}, \bibinfo {author} {\bibfnamefont {S.~D.}\ \bibnamefont
  {Odintsov}},\ and\ \bibinfo {author} {\bibfnamefont {G.~J.}\ \bibnamefont
  {Olmo}},\ }\bibfield  {title} {\bibinfo {title} {{Little Rip, $\Lambda$CDM
  and singular dark energy cosmology from Born-Infeld-$f(R)$ gravity}},\ }\href
  {https://doi.org/10.1016/j.physletb.2014.05.024} {\bibfield  {journal}
  {\bibinfo  {journal} {Phys. Lett. B}\ }\textbf {\bibinfo {volume} {734}},\
  \bibinfo {pages} {36} (\bibinfo {year} {2014}{\natexlab{b}})}\BibitemShut
  {NoStop}%
\bibitem [{\citenamefont {Odintsov}\ \emph {et~al.}(2014)\citenamefont
  {Odintsov}, \citenamefont {Olmo},\ and\ \citenamefont
  {Rubiera-Garcia}}]{odintsov2014born}%
  \BibitemOpen
  \bibfield  {author} {\bibinfo {author} {\bibfnamefont {S.~D.}\ \bibnamefont
  {Odintsov}}, \bibinfo {author} {\bibfnamefont {G.~J.}\ \bibnamefont {Olmo}},\
  and\ \bibinfo {author} {\bibfnamefont {D.}~\bibnamefont {Rubiera-Garcia}},\
  }\bibfield  {title} {\bibinfo {title} {{Born-Infeld gravity and its
  functional extensions}},\ }\href {https://doi.org/10.1103/PhysRevD.90.044003}
  {\bibfield  {journal} {\bibinfo  {journal} {Phys. Rev. D}\ }\textbf {\bibinfo
  {volume} {90}},\ \bibinfo {pages} {044003} (\bibinfo {year}
  {2014})}\BibitemShut {NoStop}%
\bibitem [{\citenamefont {Elizalde}\ and\ \citenamefont
  {Makarenko}(2016)}]{Elizalde:2016vsd}%
  \BibitemOpen
  \bibfield  {author} {\bibinfo {author} {\bibfnamefont {E.}~\bibnamefont
  {Elizalde}}\ and\ \bibinfo {author} {\bibfnamefont {A.~N.}\ \bibnamefont
  {Makarenko}},\ }\bibfield  {title} {\bibinfo {title} {{Singular inflation
  from Born\textendash{}Infeld-f (R) gravity}},\ }\href
  {https://doi.org/10.1142/S0217732316501492} {\bibfield  {journal} {\bibinfo
  {journal} {Mod. Phys. Lett. A}\ }\textbf {\bibinfo {volume} {31}},\ \bibinfo
  {pages} {1650149} (\bibinfo {year} {2016})}\BibitemShut {NoStop}%
\bibitem [{\citenamefont {Kibaro\u{g}lu}(2023)}]{Kibaroglu:2023BIdS}%
  \BibitemOpen
  \bibfield  {author} {\bibinfo {author} {\bibfnamefont {S.}~\bibnamefont
  {Kibaro\u{g}lu}},\ }\bibfield  {title} {\bibinfo {title}
  {{Born\textendash{}Infeld-$f(R)$ gravity with de Sitter solutions}},\ }\href
  {https://doi.org/10.1142/S0219887823501414} {\bibfield  {journal} {\bibinfo
  {journal} {Int. J. Geom. Meth. Mod. Phys.}\ }\textbf {\bibinfo {volume}
  {20}},\ \bibinfo {pages} {2350141} (\bibinfo {year} {2023})}\BibitemShut
  {NoStop}%
\bibitem [{\citenamefont {Kibaro\u{g}lu}\ and\ \citenamefont
  {Elizalde}(2023)}]{Kibaroglu:2023CosmBI}%
  \BibitemOpen
  \bibfield  {author} {\bibinfo {author} {\bibfnamefont {S.}~\bibnamefont
  {Kibaro\u{g}lu}}\ and\ \bibinfo {author} {\bibfnamefont {E.}~\bibnamefont
  {Elizalde}},\ }\bibfield  {title} {\bibinfo {title} {{Cosmological
  implications of Born\textendash{}Infeld-$f(R)$ gravity}},\ }\href
  {https://doi.org/10.1142/S0218271823500128} {\bibfield  {journal} {\bibinfo
  {journal} {Int. J. Mod. Phys. D}\ }\textbf {\bibinfo {volume} {32}},\
  \bibinfo {pages} {2350012} (\bibinfo {year} {2023})}\BibitemShut {NoStop}%
\bibitem [{\citenamefont {Kibaro\u{g}lu}\ \emph {et~al.}(2024)\citenamefont
  {Kibaro\u{g}lu}, \citenamefont {Odintsov},\ and\ \citenamefont
  {Paul}}]{Kibaroglu:2024UnimodularBI}%
  \BibitemOpen
  \bibfield  {author} {\bibinfo {author} {\bibfnamefont {S.}~\bibnamefont
  {Kibaro\u{g}lu}}, \bibinfo {author} {\bibfnamefont {S.~D.}\ \bibnamefont
  {Odintsov}},\ and\ \bibinfo {author} {\bibfnamefont {T.}~\bibnamefont
  {Paul}},\ }\bibfield  {title} {\bibinfo {title} {{Cosmology of unimodular
  Born\textendash{}Infeld-$f(R)$ gravity}},\ }\href
  {https://doi.org/10.1016/j.dark.2024.101445} {\bibfield  {journal} {\bibinfo
  {journal} {Phys. Dark Univ.}\ }\textbf {\bibinfo {volume} {44}},\ \bibinfo
  {pages} {101445} (\bibinfo {year} {2024})}\BibitemShut {NoStop}%
\bibitem [{\citenamefont {Beltran~Jimenez}\ \emph {et~al.}(2018)\citenamefont
  {Beltran~Jimenez}, \citenamefont {Heisenberg}, \citenamefont {Olmo},\ and\
  \citenamefont {Rubiera-Garcia}}]{jimenez2018born}%
  \BibitemOpen
  \bibfield  {author} {\bibinfo {author} {\bibfnamefont {J.}~\bibnamefont
  {Beltran~Jimenez}}, \bibinfo {author} {\bibfnamefont {L.}~\bibnamefont
  {Heisenberg}}, \bibinfo {author} {\bibfnamefont {G.~J.}\ \bibnamefont
  {Olmo}},\ and\ \bibinfo {author} {\bibfnamefont {D.}~\bibnamefont
  {Rubiera-Garcia}},\ }\bibfield  {title} {\bibinfo {title}
  {{Born\textendash{}Infeld inspired modifications of gravity}},\ }\href
  {https://doi.org/10.1016/j.physrep.2017.11.001} {\bibfield  {journal}
  {\bibinfo  {journal} {Phys. Rept.}\ }\textbf {\bibinfo {volume} {727}},\
  \bibinfo {pages} {1} (\bibinfo {year} {2018})}\BibitemShut {NoStop}%
\bibitem [{\citenamefont {Hinshaw}\ \emph {et~al.}(2003)\citenamefont {Hinshaw}
  \emph {et~al.}}]{WMAP:2003}%
  \BibitemOpen
  \bibfield  {author} {\bibinfo {author} {\bibfnamefont {G.}~\bibnamefont
  {Hinshaw}} \emph {et~al.} (\bibinfo {collaboration} {WMAP}),\ }\bibfield
  {title} {\bibinfo {title} {{First year Wilkinson Microwave Anisotropy Probe
  (WMAP) observations: The Angular power spectrum}},\ }\href
  {https://doi.org/10.1086/377225} {\bibfield  {journal} {\bibinfo  {journal}
  {Astrophys. J. Suppl.}\ }\textbf {\bibinfo {volume} {148}},\ \bibinfo {pages}
  {135} (\bibinfo {year} {2003})}\BibitemShut {NoStop}%
\bibitem [{\citenamefont {Hinshaw}\ \emph {et~al.}(2007)\citenamefont {Hinshaw}
  \emph {et~al.}}]{WMAP:2006}%
  \BibitemOpen
  \bibfield  {author} {\bibinfo {author} {\bibfnamefont {G.}~\bibnamefont
  {Hinshaw}} \emph {et~al.} (\bibinfo {collaboration} {WMAP}),\ }\bibfield
  {title} {\bibinfo {title} {{Three-year Wilkinson Microwave Anisotropy Probe
  (WMAP) observations: temperature analysis}},\ }\href
  {https://doi.org/10.1086/513698} {\bibfield  {journal} {\bibinfo  {journal}
  {Astrophys. J. Suppl.}\ }\textbf {\bibinfo {volume} {170}},\ \bibinfo {pages}
  {288} (\bibinfo {year} {2007})}\BibitemShut {NoStop}%
\bibitem [{\citenamefont {Hinshaw}\ \emph {et~al.}(2009)\citenamefont {Hinshaw}
  \emph {et~al.}}]{WMAP:2008}%
  \BibitemOpen
  \bibfield  {author} {\bibinfo {author} {\bibfnamefont {G.}~\bibnamefont
  {Hinshaw}} \emph {et~al.} (\bibinfo {collaboration} {WMAP}),\ }\bibfield
  {title} {\bibinfo {title} {{Five-Year Wilkinson Microwave Anisotropy Probe
  (WMAP) Observations: Data Processing, Sky Maps, and Basic Results}},\ }\href
  {https://doi.org/10.1088/0067-0049/180/2/225} {\bibfield  {journal} {\bibinfo
   {journal} {Astrophys. J. Suppl.}\ }\textbf {\bibinfo {volume} {180}},\
  \bibinfo {pages} {225} (\bibinfo {year} {2009})}\BibitemShut {NoStop}%
\bibitem [{\citenamefont {Jaffe}\ \emph {et~al.}(2005)\citenamefont {Jaffe},
  \citenamefont {Banday}, \citenamefont {Eriksen}, \citenamefont {Gorski},\
  and\ \citenamefont {Hansen}}]{Jaffe:2005Evidence}%
  \BibitemOpen
  \bibfield  {author} {\bibinfo {author} {\bibfnamefont {T.~R.}\ \bibnamefont
  {Jaffe}}, \bibinfo {author} {\bibfnamefont {A.~J.}\ \bibnamefont {Banday}},
  \bibinfo {author} {\bibfnamefont {H.~K.}\ \bibnamefont {Eriksen}}, \bibinfo
  {author} {\bibfnamefont {K.~M.}\ \bibnamefont {Gorski}},\ and\ \bibinfo
  {author} {\bibfnamefont {F.~K.}\ \bibnamefont {Hansen}},\ }\bibfield  {title}
  {\bibinfo {title} {{Evidence of vorticity and shear at large angular scales
  in the WMAP data: A Violation of cosmological isotropy?}},\ }\href
  {https://doi.org/10.1086/444454} {\bibfield  {journal} {\bibinfo  {journal}
  {Astrophys. J. Lett.}\ }\textbf {\bibinfo {volume} {629}},\ \bibinfo {pages}
  {L1} (\bibinfo {year} {2005})}\BibitemShut {NoStop}%
\bibitem [{\citenamefont {Jaffe}\ \emph
  {et~al.}(2006{\natexlab{a}})\citenamefont {Jaffe}, \citenamefont {Banday},
  \citenamefont {Eriksen}, \citenamefont {Gorski},\ and\ \citenamefont
  {Hansen}}]{Jaffe:2006Bianchi}%
  \BibitemOpen
  \bibfield  {author} {\bibinfo {author} {\bibfnamefont {T.~R.}\ \bibnamefont
  {Jaffe}}, \bibinfo {author} {\bibfnamefont {A.~J.}\ \bibnamefont {Banday}},
  \bibinfo {author} {\bibfnamefont {H.~K.}\ \bibnamefont {Eriksen}}, \bibinfo
  {author} {\bibfnamefont {K.~M.}\ \bibnamefont {Gorski}},\ and\ \bibinfo
  {author} {\bibfnamefont {F.~K.}\ \bibnamefont {Hansen}},\ }\bibfield  {title}
  {\bibinfo {title} {{Bianchi Type VII(h) Models and the WMAP 3-year Data}},\
  }\href {https://doi.org/10.1051/0004-6361:20065748} {\bibfield  {journal}
  {\bibinfo  {journal} {Astron. Astrophys.}\ }\textbf {\bibinfo {volume}
  {460}},\ \bibinfo {pages} {393} (\bibinfo {year}
  {2006}{\natexlab{a}})}\BibitemShut {NoStop}%
\bibitem [{\citenamefont {Jaffe}\ \emph
  {et~al.}(2006{\natexlab{b}})\citenamefont {Jaffe}, \citenamefont {Banday},
  \citenamefont {Eriksen}, \citenamefont {Gorski},\ and\ \citenamefont
  {Hansen}}]{Jaffe:2006Fast}%
  \BibitemOpen
  \bibfield  {author} {\bibinfo {author} {\bibfnamefont {T.~R.}\ \bibnamefont
  {Jaffe}}, \bibinfo {author} {\bibfnamefont {A.~J.}\ \bibnamefont {Banday}},
  \bibinfo {author} {\bibfnamefont {H.~K.}\ \bibnamefont {Eriksen}}, \bibinfo
  {author} {\bibfnamefont {K.~M.}\ \bibnamefont {Gorski}},\ and\ \bibinfo
  {author} {\bibfnamefont {F.~K.}\ \bibnamefont {Hansen}},\ }\bibfield  {title}
  {\bibinfo {title} {{Fast and efficient template fitting of deterministic
  anisotropic cosmological models applied to wmap data}},\ }\href
  {https://doi.org/10.1086/501343} {\bibfield  {journal} {\bibinfo  {journal}
  {Astrophys. J.}\ }\textbf {\bibinfo {volume} {643}},\ \bibinfo {pages} {616}
  (\bibinfo {year} {2006}{\natexlab{b}})}\BibitemShut {NoStop}%
\bibitem [{\citenamefont {Campanelli}\ \emph {et~al.}(2006)\citenamefont
  {Campanelli}, \citenamefont {Cea},\ and\ \citenamefont
  {Tedesco}}]{Campanelli:2006Ellipsoidal}%
  \BibitemOpen
  \bibfield  {author} {\bibinfo {author} {\bibfnamefont {L.}~\bibnamefont
  {Campanelli}}, \bibinfo {author} {\bibfnamefont {P.}~\bibnamefont {Cea}},\
  and\ \bibinfo {author} {\bibfnamefont {L.}~\bibnamefont {Tedesco}},\
  }\bibfield  {title} {\bibinfo {title} {{Ellipsoidal Universe Can Solve The
  CMB Quadrupole Problem}},\ }\href
  {https://doi.org/10.1103/PhysRevLett.97.131302} {\bibfield  {journal}
  {\bibinfo  {journal} {Phys. Rev. Lett.}\ }\textbf {\bibinfo {volume} {97}},\
  \bibinfo {pages} {131302} (\bibinfo {year} {2006})},\ \bibinfo {note}
  {[Erratum: Phys.Rev.Lett. 97, 209903 (2006)]}\BibitemShut {NoStop}%
\bibitem [{\citenamefont {Campanelli}\ \emph {et~al.}(2007)\citenamefont
  {Campanelli}, \citenamefont {Cea},\ and\ \citenamefont
  {Tedesco}}]{Campanelli:2007Cosmic}%
  \BibitemOpen
  \bibfield  {author} {\bibinfo {author} {\bibfnamefont {L.}~\bibnamefont
  {Campanelli}}, \bibinfo {author} {\bibfnamefont {P.}~\bibnamefont {Cea}},\
  and\ \bibinfo {author} {\bibfnamefont {L.}~\bibnamefont {Tedesco}},\
  }\bibfield  {title} {\bibinfo {title} {{Cosmic Microwave Background
  Quadrupole and Ellipsoidal Universe}},\ }\href
  {https://doi.org/10.1103/PhysRevD.76.063007} {\bibfield  {journal} {\bibinfo
  {journal} {Phys. Rev. D}\ }\textbf {\bibinfo {volume} {76}},\ \bibinfo
  {pages} {063007} (\bibinfo {year} {2007})}\BibitemShut {NoStop}%
\bibitem [{\citenamefont {Guth}(1981)}]{guth1981inflationary}%
  \BibitemOpen
  \bibfield  {author} {\bibinfo {author} {\bibfnamefont {A.~H.}\ \bibnamefont
  {Guth}},\ }\bibfield  {title} {\bibinfo {title} {{The Inflationary Universe:
  A Possible Solution to the Horizon and Flatness Problems}},\ }\href
  {https://doi.org/10.1103/PhysRevD.23.347} {\bibfield  {journal} {\bibinfo
  {journal} {Phys. Rev. D}\ }\textbf {\bibinfo {volume} {23}},\ \bibinfo
  {pages} {347} (\bibinfo {year} {1981})}\BibitemShut {NoStop}%
\bibitem [{\citenamefont {Linde}(1982)}]{linde1982new}%
  \BibitemOpen
  \bibfield  {author} {\bibinfo {author} {\bibfnamefont {A.~D.}\ \bibnamefont
  {Linde}},\ }\bibfield  {title} {\bibinfo {title} {{A New Inflationary
  Universe Scenario: A Possible Solution of the Horizon, Flatness, Homogeneity,
  Isotropy and Primordial Monopole Problems}},\ }\href
  {https://doi.org/10.1016/0370-2693(82)91219-9} {\bibfield  {journal}
  {\bibinfo  {journal} {Phys. Lett. B}\ }\textbf {\bibinfo {volume} {108}},\
  \bibinfo {pages} {389} (\bibinfo {year} {1982})}\BibitemShut {NoStop}%
\bibitem [{\citenamefont {Linde}(1983)}]{linde1983chaotic}%
  \BibitemOpen
  \bibfield  {author} {\bibinfo {author} {\bibfnamefont {A.~D.}\ \bibnamefont
  {Linde}},\ }\bibfield  {title} {\bibinfo {title} {{Chaotic Inflation}},\
  }\href {https://doi.org/10.1016/0370-2693(83)90837-7} {\bibfield  {journal}
  {\bibinfo  {journal} {Phys. Lett. B}\ }\textbf {\bibinfo {volume} {129}},\
  \bibinfo {pages} {177} (\bibinfo {year} {1983})}\BibitemShut {NoStop}%
\bibitem [{\citenamefont {Linde}(1991)}]{linde1991axions}%
  \BibitemOpen
  \bibfield  {author} {\bibinfo {author} {\bibfnamefont {A.~D.}\ \bibnamefont
  {Linde}},\ }\bibfield  {title} {\bibinfo {title} {{Axions in inflationary
  cosmology}},\ }\href {https://doi.org/10.1016/0370-2693(91)90130-I}
  {\bibfield  {journal} {\bibinfo  {journal} {Phys. Lett. B}\ }\textbf
  {\bibinfo {volume} {259}},\ \bibinfo {pages} {38} (\bibinfo {year}
  {1991})}\BibitemShut {NoStop}%
\bibitem [{\citenamefont {Linde}(1994)}]{linde1994hybrid}%
  \BibitemOpen
  \bibfield  {author} {\bibinfo {author} {\bibfnamefont {A.~D.}\ \bibnamefont
  {Linde}},\ }\bibfield  {title} {\bibinfo {title} {{Hybrid inflation}},\
  }\href {https://doi.org/10.1103/PhysRevD.49.748} {\bibfield  {journal}
  {\bibinfo  {journal} {Phys. Rev. D}\ }\textbf {\bibinfo {volume} {49}},\
  \bibinfo {pages} {748} (\bibinfo {year} {1994})}\BibitemShut {NoStop}%
\bibitem [{\citenamefont {Collins}\ \emph {et~al.}(1980)\citenamefont
  {Collins}, \citenamefont {Glass},\ and\ \citenamefont
  {Wilkinson}}]{Collins:1980Exact}%
  \BibitemOpen
  \bibfield  {author} {\bibinfo {author} {\bibfnamefont {C.~B.}\ \bibnamefont
  {Collins}}, \bibinfo {author} {\bibfnamefont {E.~N.}\ \bibnamefont {Glass}},\
  and\ \bibinfo {author} {\bibfnamefont {D.~A.}\ \bibnamefont {Wilkinson}},\
  }\bibfield  {title} {\bibinfo {title} {{Exact spatially homogeneous
  cosmologies}},\ }\href {https://doi.org/10.1007/BF00763057} {\bibfield
  {journal} {\bibinfo  {journal} {Gen. Rel. Grav.}\ }\textbf {\bibinfo {volume}
  {12}},\ \bibinfo {pages} {805} (\bibinfo {year} {1980})}\BibitemShut
  {NoStop}%
\bibitem [{\citenamefont {Ellis}(2006)}]{Ellis:2006TheBianchiModels}%
  \BibitemOpen
  \bibfield  {author} {\bibinfo {author} {\bibfnamefont {G.~F.~R.}\
  \bibnamefont {Ellis}},\ }\bibfield  {title} {\bibinfo {title} {{The Bianchi
  models: Then and now}},\ }\href {https://doi.org/10.1007/s10714-006-0283-4}
  {\bibfield  {journal} {\bibinfo  {journal} {Gen. Rel. Grav.}\ }\textbf
  {\bibinfo {volume} {38}},\ \bibinfo {pages} {1003} (\bibinfo {year}
  {2006})}\BibitemShut {NoStop}%
\bibitem [{\citenamefont {Gumrukcuoglu}\ \emph {et~al.}(2007)\citenamefont
  {Gumrukcuoglu}, \citenamefont {Contaldi},\ and\ \citenamefont
  {Peloso}}]{Gumrukcuoglu:2007Inflationary}%
  \BibitemOpen
  \bibfield  {author} {\bibinfo {author} {\bibfnamefont {A.~E.}\ \bibnamefont
  {Gumrukcuoglu}}, \bibinfo {author} {\bibfnamefont {C.~R.}\ \bibnamefont
  {Contaldi}},\ and\ \bibinfo {author} {\bibfnamefont {M.}~\bibnamefont
  {Peloso}},\ }\bibfield  {title} {\bibinfo {title} {{Inflationary
  perturbations in anisotropic backgrounds and their imprint on the CMB}},\
  }\href {https://doi.org/10.1088/1475-7516/2007/11/005} {\bibfield  {journal}
  {\bibinfo  {journal} {JCAP}\ }\textbf {\bibinfo {volume} {11}},\ \bibinfo
  {pages} {005}}\BibitemShut {NoStop}%
\bibitem [{\citenamefont {Akarsu}\ and\ \citenamefont
  {Kilinc}(2010)}]{Akarsu:2010BianchiTypeIII}%
  \BibitemOpen
  \bibfield  {author} {\bibinfo {author} {\bibfnamefont {O.}~\bibnamefont
  {Akarsu}}\ and\ \bibinfo {author} {\bibfnamefont {C.~B.}\ \bibnamefont
  {Kilinc}},\ }\bibfield  {title} {\bibinfo {title} {{Bianchi type III models
  with anisotropic dark energy}},\ }\href
  {https://doi.org/10.1007/s10714-009-0878-7} {\bibfield  {journal} {\bibinfo
  {journal} {Gen. Rel. Grav.}\ }\textbf {\bibinfo {volume} {42}},\ \bibinfo
  {pages} {763} (\bibinfo {year} {2010})}\BibitemShut {NoStop}%
\bibitem [{\citenamefont {M\"uller}\ \emph {et~al.}(2018)\citenamefont
  {M\"uller}, \citenamefont {Ricciardone}, \citenamefont {Starobinsky},\ and\
  \citenamefont {Toporensky}}]{Muller:2018Anisotropic}%
  \BibitemOpen
  \bibfield  {author} {\bibinfo {author} {\bibfnamefont {D.}~\bibnamefont
  {M\"uller}}, \bibinfo {author} {\bibfnamefont {A.}~\bibnamefont
  {Ricciardone}}, \bibinfo {author} {\bibfnamefont {A.~A.}\ \bibnamefont
  {Starobinsky}},\ and\ \bibinfo {author} {\bibfnamefont {A.}~\bibnamefont
  {Toporensky}},\ }\bibfield  {title} {\bibinfo {title} {{Anisotropic
  cosmological solutions in $R + R^2$ gravity}},\ }\href
  {https://doi.org/10.1140/epjc/s10052-018-5778-0} {\bibfield  {journal}
  {\bibinfo  {journal} {Eur. Phys. J. C}\ }\textbf {\bibinfo {volume} {78}},\
  \bibinfo {pages} {311} (\bibinfo {year} {2018})},\ \Eprint
  {https://arxiv.org/abs/1710.08753} {arXiv:1710.08753 [gr-qc]} \BibitemShut
  {NoStop}%
\bibitem [{\citenamefont {Amirhashchi}\ and\ \citenamefont
  {Amirhashchi}(2020)}]{Amirhashchi:2020Constraining}%
  \BibitemOpen
  \bibfield  {author} {\bibinfo {author} {\bibfnamefont {H.}~\bibnamefont
  {Amirhashchi}}\ and\ \bibinfo {author} {\bibfnamefont {S.}~\bibnamefont
  {Amirhashchi}},\ }\bibfield  {title} {\bibinfo {title} {{Constraining Bianchi
  Type I Universe With Type Ia Supernova and H(z) Data}},\ }\href
  {https://doi.org/10.1016/j.dark.2020.100557} {\bibfield  {journal} {\bibinfo
  {journal} {Phys. Dark Univ.}\ }\textbf {\bibinfo {volume} {29}},\ \bibinfo
  {pages} {100557} (\bibinfo {year} {2020})}\BibitemShut {NoStop}%
\bibitem [{\citenamefont {Kumar}\ \emph {et~al.}(2021)\citenamefont {Kumar},
  \citenamefont {Maheshwari}, \citenamefont {Mazumdar},\ and\ \citenamefont
  {Peng}}]{Kumar:2021Anisotropic}%
  \BibitemOpen
  \bibfield  {author} {\bibinfo {author} {\bibfnamefont {K.~S.}\ \bibnamefont
  {Kumar}}, \bibinfo {author} {\bibfnamefont {S.}~\bibnamefont {Maheshwari}},
  \bibinfo {author} {\bibfnamefont {A.}~\bibnamefont {Mazumdar}},\ and\
  \bibinfo {author} {\bibfnamefont {J.}~\bibnamefont {Peng}},\ }\bibfield
  {title} {\bibinfo {title} {{An anisotropic bouncing universe in non-local
  gravity}},\ }\href {https://doi.org/10.1088/1475-7516/2021/07/025} {\bibfield
   {journal} {\bibinfo  {journal} {JCAP}\ }\textbf {\bibinfo {volume} {07}},\
  \bibinfo {pages} {025}}\BibitemShut {NoStop}%
\bibitem [{\citenamefont {Costantini}\ and\ \citenamefont
  {Elizalde}(2022)}]{Costantini:2022AReconstruction}%
  \BibitemOpen
  \bibfield  {author} {\bibinfo {author} {\bibfnamefont {A.}~\bibnamefont
  {Costantini}}\ and\ \bibinfo {author} {\bibfnamefont {E.}~\bibnamefont
  {Elizalde}},\ }\bibfield  {title} {\bibinfo {title} {{A reconstruction method
  for anisotropic universes in unimodular F(R)-gravity}},\ }\href
  {https://doi.org/10.1140/epjc/s10052-022-11112-3} {\bibfield  {journal}
  {\bibinfo  {journal} {Eur. Phys. J. C}\ }\textbf {\bibinfo {volume} {82}},\
  \bibinfo {pages} {1127} (\bibinfo {year} {2022})}\BibitemShut {NoStop}%
\bibitem [{\citenamefont {Nojiri}\ \emph {et~al.}(2022)\citenamefont {Nojiri},
  \citenamefont {Odintsov}, \citenamefont {Oikonomou},\ and\ \citenamefont
  {Constantini}}]{Nojiri:2022Formalizing}%
  \BibitemOpen
  \bibfield  {author} {\bibinfo {author} {\bibfnamefont {S.}~\bibnamefont
  {Nojiri}}, \bibinfo {author} {\bibfnamefont {S.~D.}\ \bibnamefont
  {Odintsov}}, \bibinfo {author} {\bibfnamefont {V.~K.}\ \bibnamefont
  {Oikonomou}},\ and\ \bibinfo {author} {\bibfnamefont {A.}~\bibnamefont
  {Constantini}},\ }\bibfield  {title} {\bibinfo {title} {{Formalizing
  anisotropic inflation in modified gravity}},\ }\href
  {https://doi.org/10.1016/j.nuclphysb.2022.116011} {\bibfield  {journal}
  {\bibinfo  {journal} {Nucl. Phys. B}\ }\textbf {\bibinfo {volume} {985}},\
  \bibinfo {pages} {116011} (\bibinfo {year} {2022})}\BibitemShut {NoStop}%
\bibitem [{\citenamefont {Parnovsky}(2023)}]{Parnovsky:2023TheBigBang}%
  \BibitemOpen
  \bibfield  {author} {\bibinfo {author} {\bibfnamefont {S.~L.}\ \bibnamefont
  {Parnovsky}},\ }\bibfield  {title} {\bibinfo {title} {{The Big Bang could be
  anisotropic. The case of Bianchi I model}},\ }\href
  {https://doi.org/10.1088/1361-6382/acd7c2} {\bibfield  {journal} {\bibinfo
  {journal} {Class. Quant. Grav.}\ }\textbf {\bibinfo {volume} {40}},\ \bibinfo
  {pages} {135005} (\bibinfo {year} {2023})}\BibitemShut {NoStop}%
\bibitem [{\citenamefont {Vollick}(2003)}]{Vollick:2003AnisotropicBI}%
  \BibitemOpen
  \bibfield  {author} {\bibinfo {author} {\bibfnamefont {D.~N.}\ \bibnamefont
  {Vollick}},\ }\bibfield  {title} {\bibinfo {title} {{Anisotropic Born-Infeld
  cosmologies}},\ }\href {https://doi.org/10.1023/A:1024551105800} {\bibfield
  {journal} {\bibinfo  {journal} {Gen. Rel. Grav.}\ }\textbf {\bibinfo {volume}
  {35}},\ \bibinfo {pages} {1511} (\bibinfo {year} {2003})}\BibitemShut
  {NoStop}%
\bibitem [{\citenamefont {Rodrigues}(2008)}]{Rodrigues:2008Evolution}%
  \BibitemOpen
  \bibfield  {author} {\bibinfo {author} {\bibfnamefont {D.~C.}\ \bibnamefont
  {Rodrigues}},\ }\bibfield  {title} {\bibinfo {title} {{Evolution of
  Anisotropies in Eddington-Born-Infeld Cosmology}},\ }\href
  {https://doi.org/10.1103/PhysRevD.78.063013} {\bibfield  {journal} {\bibinfo
  {journal} {Phys. Rev. D}\ }\textbf {\bibinfo {volume} {78}},\ \bibinfo
  {pages} {063013} (\bibinfo {year} {2008})}\BibitemShut {NoStop}%
\bibitem [{\citenamefont {Harko}\ \emph {et~al.}(2014)\citenamefont {Harko},
  \citenamefont {Lobo},\ and\ \citenamefont {Mak}}]{Harko:2014Bianchi}%
  \BibitemOpen
  \bibfield  {author} {\bibinfo {author} {\bibfnamefont {T.}~\bibnamefont
  {Harko}}, \bibinfo {author} {\bibfnamefont {F.~S.~N.}\ \bibnamefont {Lobo}},\
  and\ \bibinfo {author} {\bibfnamefont {M.~K.}\ \bibnamefont {Mak}},\
  }\bibfield  {title} {\bibinfo {title} {{Bianchi type I cosmological models in
  Eddington-inspired Born-Infeld gravity}},\ }\href
  {https://doi.org/10.3390/galaxies2040496} {\bibfield  {journal} {\bibinfo
  {journal} {Galaxies}\ }\textbf {\bibinfo {volume} {2}},\ \bibinfo {pages}
  {496} (\bibinfo {year} {2014})}\BibitemShut {NoStop}%
\bibitem [{\citenamefont {Olmo}(2011)}]{Olmo:2011uz}%
  \BibitemOpen
  \bibfield  {author} {\bibinfo {author} {\bibfnamefont {G.~J.}\ \bibnamefont
  {Olmo}},\ }\bibfield  {title} {\bibinfo {title} {{Palatini Approach to
  Modified Gravity: f(R) Theories and Beyond}},\ }\href
  {https://doi.org/10.1142/S0218271811018925} {\bibfield  {journal} {\bibinfo
  {journal} {Int. J. Mod. Phys. D}\ }\textbf {\bibinfo {volume} {20}},\
  \bibinfo {pages} {413} (\bibinfo {year} {2011})}\BibitemShut {NoStop}%
\bibitem [{\citenamefont {Olmo}\ and\ \citenamefont
  {Olmo}(2012)}]{olmo2012open}%
  \BibitemOpen
  \bibinfo {editor} {\bibfnamefont {G.~J.}\ \bibnamefont {Olmo}}\ and\ \bibinfo
  {editor} {\bibfnamefont {G.~J.}\ \bibnamefont {Olmo}},\ eds.,\ \href
  {https://doi.org/10.5772/45746} {\emph {\bibinfo {title} {{Open Questions in
  Cosmology}}}}\ (\bibinfo  {publisher} {InTech},\ \bibinfo {year}
  {2012})\BibitemShut {NoStop}%
\bibitem [{\citenamefont {Nojiri}\ and\ \citenamefont
  {Odintsov}(2006)}]{Nojiri:2006Modified}%
  \BibitemOpen
  \bibfield  {author} {\bibinfo {author} {\bibfnamefont {S.}~\bibnamefont
  {Nojiri}}\ and\ \bibinfo {author} {\bibfnamefont {S.~D.}\ \bibnamefont
  {Odintsov}},\ }\bibfield  {title} {\bibinfo {title} {{Modified f(R) gravity
  consistent with realistic cosmology: From matter dominated epoch to dark
  energy universe}},\ }\href {https://doi.org/10.1103/PhysRevD.74.086005}
  {\bibfield  {journal} {\bibinfo  {journal} {Phys. Rev. D}\ }\textbf {\bibinfo
  {volume} {74}},\ \bibinfo {pages} {086005} (\bibinfo {year}
  {2006})}\BibitemShut {NoStop}%
\bibitem [{\citenamefont {Cognola}\ \emph {et~al.}(2007)\citenamefont
  {Cognola}, \citenamefont {Elizalde}, \citenamefont {Nojiri}, \citenamefont
  {Odintsov},\ and\ \citenamefont {Zerbini}}]{Cognola:2007String}%
  \BibitemOpen
  \bibfield  {author} {\bibinfo {author} {\bibfnamefont {G.}~\bibnamefont
  {Cognola}}, \bibinfo {author} {\bibfnamefont {E.}~\bibnamefont {Elizalde}},
  \bibinfo {author} {\bibfnamefont {S.}~\bibnamefont {Nojiri}}, \bibinfo
  {author} {\bibfnamefont {S.}~\bibnamefont {Odintsov}},\ and\ \bibinfo
  {author} {\bibfnamefont {S.}~\bibnamefont {Zerbini}},\ }\bibfield  {title}
  {\bibinfo {title} {{String-inspired Gauss-Bonnet gravity reconstructed from
  the universe expansion history and yielding the transition from matter
  dominance to dark energy}},\ }\href
  {https://doi.org/10.1103/PhysRevD.75.086002} {\bibfield  {journal} {\bibinfo
  {journal} {Phys. Rev. D}\ }\textbf {\bibinfo {volume} {75}},\ \bibinfo
  {pages} {086002} (\bibinfo {year} {2007})}\BibitemShut {NoStop}%
\bibitem [{\citenamefont {Cognola}\ \emph {et~al.}(2008)\citenamefont
  {Cognola}, \citenamefont {Elizalde}, \citenamefont {Nojiri}, \citenamefont
  {Odintsov}, \citenamefont {Sebastiani},\ and\ \citenamefont
  {Zerbini}}]{Cognola:2008AClass}%
  \BibitemOpen
  \bibfield  {author} {\bibinfo {author} {\bibfnamefont {G.}~\bibnamefont
  {Cognola}}, \bibinfo {author} {\bibfnamefont {E.}~\bibnamefont {Elizalde}},
  \bibinfo {author} {\bibfnamefont {S.}~\bibnamefont {Nojiri}}, \bibinfo
  {author} {\bibfnamefont {S.~D.}\ \bibnamefont {Odintsov}}, \bibinfo {author}
  {\bibfnamefont {L.}~\bibnamefont {Sebastiani}},\ and\ \bibinfo {author}
  {\bibfnamefont {S.}~\bibnamefont {Zerbini}},\ }\bibfield  {title} {\bibinfo
  {title} {{A Class of viable modified f(R) gravities describing inflation and
  the onset of accelerated expansion}},\ }\href
  {https://doi.org/10.1103/PhysRevD.77.046009} {\bibfield  {journal} {\bibinfo
  {journal} {Phys. Rev. D}\ }\textbf {\bibinfo {volume} {77}},\ \bibinfo
  {pages} {046009} (\bibinfo {year} {2008})}\BibitemShut {NoStop}%
\bibitem [{\citenamefont {Elizalde}\ \emph {et~al.}(2008)\citenamefont
  {Elizalde}, \citenamefont {Nojiri}, \citenamefont {Odintsov}, \citenamefont
  {Saez-Gomez},\ and\ \citenamefont {Faraoni}}]{Elizalde:2008Reconstructing}%
  \BibitemOpen
  \bibfield  {author} {\bibinfo {author} {\bibfnamefont {E.}~\bibnamefont
  {Elizalde}}, \bibinfo {author} {\bibfnamefont {S.}~\bibnamefont {Nojiri}},
  \bibinfo {author} {\bibfnamefont {S.~D.}\ \bibnamefont {Odintsov}}, \bibinfo
  {author} {\bibfnamefont {D.}~\bibnamefont {Saez-Gomez}},\ and\ \bibinfo
  {author} {\bibfnamefont {V.}~\bibnamefont {Faraoni}},\ }\bibfield  {title}
  {\bibinfo {title} {{Reconstructing the universe history, from inflation to
  acceleration, with phantom and canonical scalar fields}},\ }\href
  {https://doi.org/10.1103/PhysRevD.77.106005} {\bibfield  {journal} {\bibinfo
  {journal} {Phys. Rev. D}\ }\textbf {\bibinfo {volume} {77}},\ \bibinfo
  {pages} {106005} (\bibinfo {year} {2008})}\BibitemShut {NoStop}%
\bibitem [{\citenamefont {Nojiri}\ \emph {et~al.}(2009)\citenamefont {Nojiri},
  \citenamefont {Odintsov},\ and\ \citenamefont
  {Saez-Gomez}}]{Nojiri:2009CosmologicalReconstruction}%
  \BibitemOpen
  \bibfield  {author} {\bibinfo {author} {\bibfnamefont {S.}~\bibnamefont
  {Nojiri}}, \bibinfo {author} {\bibfnamefont {S.~D.}\ \bibnamefont
  {Odintsov}},\ and\ \bibinfo {author} {\bibfnamefont {D.}~\bibnamefont
  {Saez-Gomez}},\ }\bibfield  {title} {\bibinfo {title} {{Cosmological
  reconstruction of realistic modified F(R) gravities}},\ }\href
  {https://doi.org/10.1016/j.physletb.2009.09.045} {\bibfield  {journal}
  {\bibinfo  {journal} {Phys. Lett. B}\ }\textbf {\bibinfo {volume} {681}},\
  \bibinfo {pages} {74} (\bibinfo {year} {2009})}\BibitemShut {NoStop}%
\bibitem [{\citenamefont {Taub}(1951)}]{Taub:1951Empty}%
  \BibitemOpen
  \bibfield  {author} {\bibinfo {author} {\bibfnamefont {A.~H.}\ \bibnamefont
  {Taub}},\ }\bibfield  {title} {\bibinfo {title} {{Empty space-times admitting
  a three parameter group of motions}},\ }\href
  {https://doi.org/10.2307/1969567} {\bibfield  {journal} {\bibinfo  {journal}
  {Annals Math.}\ }\textbf {\bibinfo {volume} {53}},\ \bibinfo {pages} {472}
  (\bibinfo {year} {1951})}\BibitemShut {NoStop}%
\bibitem [{\citenamefont {Barrow}\ and\ \citenamefont
  {Clifton}(2006)}]{Barrow:2006Cosmologies}%
  \BibitemOpen
  \bibfield  {author} {\bibinfo {author} {\bibfnamefont {J.~D.}\ \bibnamefont
  {Barrow}}\ and\ \bibinfo {author} {\bibfnamefont {T.}~\bibnamefont
  {Clifton}},\ }\bibfield  {title} {\bibinfo {title} {{Cosmologies with energy
  exchange}},\ }\href {https://doi.org/10.1103/PhysRevD.73.103520} {\bibfield
  {journal} {\bibinfo  {journal} {Phys. Rev. D}\ }\textbf {\bibinfo {volume}
  {73}},\ \bibinfo {pages} {103520} (\bibinfo {year} {2006})}\BibitemShut
  {NoStop}%
\bibitem [{\citenamefont {Berera}\ \emph {et~al.}(2004)\citenamefont {Berera},
  \citenamefont {Buniy},\ and\ \citenamefont
  {Kephart}}]{Berera:2004TheEccentric}%
  \BibitemOpen
  \bibfield  {author} {\bibinfo {author} {\bibfnamefont {A.}~\bibnamefont
  {Berera}}, \bibinfo {author} {\bibfnamefont {R.~V.}\ \bibnamefont {Buniy}},\
  and\ \bibinfo {author} {\bibfnamefont {T.~W.}\ \bibnamefont {Kephart}},\
  }\bibfield  {title} {\bibinfo {title} {{The Eccentric universe}},\ }\href
  {https://doi.org/10.1088/1475-7516/2004/10/016} {\bibfield  {journal}
  {\bibinfo  {journal} {JCAP}\ }\textbf {\bibinfo {volume} {10}},\ \bibinfo
  {pages} {016}}\BibitemShut {NoStop}%
\bibitem [{\citenamefont {Campanelli}(2009)}]{Campanelli:2009AModel}%
  \BibitemOpen
  \bibfield  {author} {\bibinfo {author} {\bibfnamefont {L.}~\bibnamefont
  {Campanelli}},\ }\bibfield  {title} {\bibinfo {title} {{A Model of Universe
  Anisotropization}},\ }\href {https://doi.org/10.1103/PhysRevD.80.063006}
  {\bibfield  {journal} {\bibinfo  {journal} {Phys. Rev. D}\ }\textbf {\bibinfo
  {volume} {80}},\ \bibinfo {pages} {063006} (\bibinfo {year}
  {2009})}\BibitemShut {NoStop}%
\bibitem [{\citenamefont {Starobinsky}(1980)}]{Starobinsky:1980ANew}%
  \BibitemOpen
  \bibfield  {author} {\bibinfo {author} {\bibfnamefont {A.~A.}\ \bibnamefont
  {Starobinsky}},\ }\bibfield  {title} {\bibinfo {title} {{A New Type of
  Isotropic Cosmological Models Without Singularity}},\ }\href
  {https://doi.org/10.1016/0370-2693(80)90670-X} {\bibfield  {journal}
  {\bibinfo  {journal} {Phys. Lett. B}\ }\textbf {\bibinfo {volume} {91}},\
  \bibinfo {pages} {99} (\bibinfo {year} {1980})}\BibitemShut {NoStop}%
\bibitem [{\citenamefont {Barrow}\ and\ \citenamefont
  {Cotsakis}(1988)}]{Barrow:1988Inflation}%
  \BibitemOpen
  \bibfield  {author} {\bibinfo {author} {\bibfnamefont {J.~D.}\ \bibnamefont
  {Barrow}}\ and\ \bibinfo {author} {\bibfnamefont {S.}~\bibnamefont
  {Cotsakis}},\ }\bibfield  {title} {\bibinfo {title} {{Inflation and the
  Conformal Structure of Higher Order Gravity Theories}},\ }\href
  {https://doi.org/10.1016/0370-2693(88)90110-4} {\bibfield  {journal}
  {\bibinfo  {journal} {Phys. Lett. B}\ }\textbf {\bibinfo {volume} {214}},\
  \bibinfo {pages} {515} (\bibinfo {year} {1988})}\BibitemShut {NoStop}%
\bibitem [{\citenamefont {Odintsov}\ and\ \citenamefont
  {Oikonomou}(2015)}]{Odintsov:2015SingularInflationary}%
  \BibitemOpen
  \bibfield  {author} {\bibinfo {author} {\bibfnamefont {S.~D.}\ \bibnamefont
  {Odintsov}}\ and\ \bibinfo {author} {\bibfnamefont {V.~K.}\ \bibnamefont
  {Oikonomou}},\ }\bibfield  {title} {\bibinfo {title} {{Singular Inflationary
  Universe from $F(R)$ Gravity}},\ }\href
  {https://doi.org/10.1103/PhysRevD.92.124024} {\bibfield  {journal} {\bibinfo
  {journal} {Phys. Rev. D}\ }\textbf {\bibinfo {volume} {92}},\ \bibinfo
  {pages} {124024} (\bibinfo {year} {2015})}\BibitemShut {NoStop}%
\bibitem [{\citenamefont {Akrami}\ \emph {et~al.}(2020)\citenamefont {Akrami}
  \emph {et~al.}}]{Planck:2020Planck}%
  \BibitemOpen
  \bibfield  {author} {\bibinfo {author} {\bibfnamefont {Y.}~\bibnamefont
  {Akrami}} \emph {et~al.} (\bibinfo {collaboration} {Planck}),\ }\bibfield
  {title} {\bibinfo {title} {{Planck 2018 results. X. Constraints on
  inflation}},\ }\href {https://doi.org/10.1051/0004-6361/201833887} {\bibfield
   {journal} {\bibinfo  {journal} {Astron. Astrophys.}\ }\textbf {\bibinfo
  {volume} {641}},\ \bibinfo {pages} {A10} (\bibinfo {year}
  {2020})}\BibitemShut {NoStop}%
\bibitem [{\citenamefont {Caldwell}\ \emph {et~al.}(2003)\citenamefont
  {Caldwell}, \citenamefont {Kamionkowski},\ and\ \citenamefont
  {Weinberg}}]{Caldwell:2003vq}%
  \BibitemOpen
  \bibfield  {author} {\bibinfo {author} {\bibfnamefont {R.~R.}\ \bibnamefont
  {Caldwell}}, \bibinfo {author} {\bibfnamefont {M.}~\bibnamefont
  {Kamionkowski}},\ and\ \bibinfo {author} {\bibfnamefont {N.~N.}\ \bibnamefont
  {Weinberg}},\ }\bibfield  {title} {\bibinfo {title} {{Phantom energy and
  cosmic doomsday}},\ }\href {https://doi.org/10.1103/PhysRevLett.91.071301}
  {\bibfield  {journal} {\bibinfo  {journal} {Phys. Rev. Lett.}\ }\textbf
  {\bibinfo {volume} {91}},\ \bibinfo {pages} {071301} (\bibinfo {year}
  {2003})}\BibitemShut {NoStop}%
\bibitem [{\citenamefont {Nojiri}\ and\ \citenamefont
  {Odintsov}(2003)}]{Nojiri:2003vn}%
  \BibitemOpen
  \bibfield  {author} {\bibinfo {author} {\bibfnamefont {S.}~\bibnamefont
  {Nojiri}}\ and\ \bibinfo {author} {\bibfnamefont {S.~D.}\ \bibnamefont
  {Odintsov}},\ }\bibfield  {title} {\bibinfo {title} {{Quantum de Sitter
  cosmology and phantom matter}},\ }\href
  {https://doi.org/10.1016/S0370-2693(03)00594-X} {\bibfield  {journal}
  {\bibinfo  {journal} {Phys. Lett. B}\ }\textbf {\bibinfo {volume} {562}},\
  \bibinfo {pages} {147} (\bibinfo {year} {2003})}\BibitemShut {NoStop}%
\bibitem [{\citenamefont {Faraoni}(2002)}]{Faraoni:2001tq}%
  \BibitemOpen
  \bibfield  {author} {\bibinfo {author} {\bibfnamefont {V.}~\bibnamefont
  {Faraoni}},\ }\bibfield  {title} {\bibinfo {title} {{Superquintessence}},\
  }\href {https://doi.org/10.1142/S0218271802001809} {\bibfield  {journal}
  {\bibinfo  {journal} {Int. J. Mod. Phys. D}\ }\textbf {\bibinfo {volume}
  {11}},\ \bibinfo {pages} {471} (\bibinfo {year} {2002})}\BibitemShut
  {NoStop}%
\bibitem [{\citenamefont {Frampton}\ \emph {et~al.}(2011)\citenamefont
  {Frampton}, \citenamefont {Ludwick},\ and\ \citenamefont
  {Scherrer}}]{frampton2011little}%
  \BibitemOpen
  \bibfield  {author} {\bibinfo {author} {\bibfnamefont {P.~H.}\ \bibnamefont
  {Frampton}}, \bibinfo {author} {\bibfnamefont {K.~J.}\ \bibnamefont
  {Ludwick}},\ and\ \bibinfo {author} {\bibfnamefont {R.~J.}\ \bibnamefont
  {Scherrer}},\ }\bibfield  {title} {\bibinfo {title} {{The Little Rip}},\
  }\href {https://doi.org/10.1103/PhysRevD.84.063003} {\bibfield  {journal}
  {\bibinfo  {journal} {Phys. Rev. D}\ }\textbf {\bibinfo {volume} {84}},\
  \bibinfo {pages} {063003} (\bibinfo {year} {2011})}\BibitemShut {NoStop}%
\bibitem [{\citenamefont {Frampton}\ \emph {et~al.}(2012)\citenamefont
  {Frampton}, \citenamefont {Ludwick}, \citenamefont {Nojiri}, \citenamefont
  {Odintsov},\ and\ \citenamefont {Scherrer}}]{frampton2012models}%
  \BibitemOpen
  \bibfield  {author} {\bibinfo {author} {\bibfnamefont {P.~H.}\ \bibnamefont
  {Frampton}}, \bibinfo {author} {\bibfnamefont {K.~J.}\ \bibnamefont
  {Ludwick}}, \bibinfo {author} {\bibfnamefont {S.}~\bibnamefont {Nojiri}},
  \bibinfo {author} {\bibfnamefont {S.~D.}\ \bibnamefont {Odintsov}},\ and\
  \bibinfo {author} {\bibfnamefont {R.~J.}\ \bibnamefont {Scherrer}},\
  }\bibfield  {title} {\bibinfo {title} {{Models for Little Rip Dark Energy}},\
  }\href {https://doi.org/10.1016/j.physletb.2012.01.048} {\bibfield  {journal}
  {\bibinfo  {journal} {Phys. Lett. B}\ }\textbf {\bibinfo {volume} {708}},\
  \bibinfo {pages} {204} (\bibinfo {year} {2012})}\BibitemShut {NoStop}%
\bibitem [{\citenamefont {Brevik}\ \emph {et~al.}(2011)\citenamefont {Brevik},
  \citenamefont {Elizalde}, \citenamefont {Nojiri},\ and\ \citenamefont
  {Odintsov}}]{brevik2011viscous}%
  \BibitemOpen
  \bibfield  {author} {\bibinfo {author} {\bibfnamefont {I.}~\bibnamefont
  {Brevik}}, \bibinfo {author} {\bibfnamefont {E.}~\bibnamefont {Elizalde}},
  \bibinfo {author} {\bibfnamefont {S.}~\bibnamefont {Nojiri}},\ and\ \bibinfo
  {author} {\bibfnamefont {S.~D.}\ \bibnamefont {Odintsov}},\ }\bibfield
  {title} {\bibinfo {title} {{Viscous Little Rip Cosmology}},\ }\href
  {https://doi.org/10.1103/PhysRevD.84.103508} {\bibfield  {journal} {\bibinfo
  {journal} {Phys. Rev. D}\ }\textbf {\bibinfo {volume} {84}},\ \bibinfo
  {pages} {103508} (\bibinfo {year} {2011})}\BibitemShut {NoStop}%
\bibitem [{\citenamefont {Brevik}\ \emph
  {et~al.}(2012{\natexlab{a}})\citenamefont {Brevik}, \citenamefont
  {Myrzakulov}, \citenamefont {Nojiri},\ and\ \citenamefont
  {Odintsov}}]{brevik2012turbulence}%
  \BibitemOpen
  \bibfield  {author} {\bibinfo {author} {\bibfnamefont {I.}~\bibnamefont
  {Brevik}}, \bibinfo {author} {\bibfnamefont {R.}~\bibnamefont {Myrzakulov}},
  \bibinfo {author} {\bibfnamefont {S.}~\bibnamefont {Nojiri}},\ and\ \bibinfo
  {author} {\bibfnamefont {S.~D.}\ \bibnamefont {Odintsov}},\ }\bibfield
  {title} {\bibinfo {title} {{Turbulence and Little Rip Cosmology}},\ }\href
  {https://doi.org/10.1103/PhysRevD.86.063007} {\bibfield  {journal} {\bibinfo
  {journal} {Phys. Rev. D}\ }\textbf {\bibinfo {volume} {86}},\ \bibinfo
  {pages} {063007} (\bibinfo {year} {2012}{\natexlab{a}})}\BibitemShut
  {NoStop}%
\bibitem [{\citenamefont {Brevik}\ \emph
  {et~al.}(2012{\natexlab{b}})\citenamefont {Brevik}, \citenamefont {Obukhov},
  \citenamefont {Osetrin},\ and\ \citenamefont {Timoshkin}}]{brevik2012little}%
  \BibitemOpen
  \bibfield  {author} {\bibinfo {author} {\bibfnamefont {I.}~\bibnamefont
  {Brevik}}, \bibinfo {author} {\bibfnamefont {V.~V.}\ \bibnamefont {Obukhov}},
  \bibinfo {author} {\bibfnamefont {K.~E.}\ \bibnamefont {Osetrin}},\ and\
  \bibinfo {author} {\bibfnamefont {A.~V.}\ \bibnamefont {Timoshkin}},\
  }\bibfield  {title} {\bibinfo {title} {{Little Rip cosmological models with
  time-dependent equation of state}},\ }\href
  {https://doi.org/10.1142/S0217732312502100} {\bibfield  {journal} {\bibinfo
  {journal} {Mod. Phys. Lett. A}\ }\textbf {\bibinfo {volume} {27}},\ \bibinfo
  {pages} {1250210} (\bibinfo {year} {2012}{\natexlab{b}})}\BibitemShut
  {NoStop}%
\bibitem [{\citenamefont {Makarenko}\ \emph {et~al.}(2013)\citenamefont
  {Makarenko}, \citenamefont {Obukhov},\ and\ \citenamefont
  {Kirnos}}]{makarenko2013big}%
  \BibitemOpen
  \bibfield  {author} {\bibinfo {author} {\bibfnamefont {A.~N.}\ \bibnamefont
  {Makarenko}}, \bibinfo {author} {\bibfnamefont {V.~V.}\ \bibnamefont
  {Obukhov}},\ and\ \bibinfo {author} {\bibfnamefont {I.~V.}\ \bibnamefont
  {Kirnos}},\ }\bibfield  {title} {\bibinfo {title} {{From Big to Little Rip in
  modified F(R,G) gravity}},\ }\href
  {https://doi.org/10.1007/s10509-012-1240-1} {\bibfield  {journal} {\bibinfo
  {journal} {Astrophys. Space Sci.}\ }\textbf {\bibinfo {volume} {343}},\
  \bibinfo {pages} {481} (\bibinfo {year} {2013})}\BibitemShut {NoStop}%
\bibitem [{\citenamefont {Peter}\ and\ \citenamefont
  {Pinto-Neto}(2002)}]{Peter:2002Primordial}%
  \BibitemOpen
  \bibfield  {author} {\bibinfo {author} {\bibfnamefont {P.}~\bibnamefont
  {Peter}}\ and\ \bibinfo {author} {\bibfnamefont {N.}~\bibnamefont
  {Pinto-Neto}},\ }\bibfield  {title} {\bibinfo {title} {{Primordial
  perturbations in a non singular bouncing universe model}},\ }\href
  {https://doi.org/10.1103/PhysRevD.66.063509} {\bibfield  {journal} {\bibinfo
  {journal} {Phys. Rev. D}\ }\textbf {\bibinfo {volume} {66}},\ \bibinfo
  {pages} {063509} (\bibinfo {year} {2002})}\BibitemShut {NoStop}%
\bibitem [{\citenamefont {Allen}\ and\ \citenamefont
  {Wands}(2004)}]{Allen:2004Cosmological}%
  \BibitemOpen
  \bibfield  {author} {\bibinfo {author} {\bibfnamefont {L.~E.}\ \bibnamefont
  {Allen}}\ and\ \bibinfo {author} {\bibfnamefont {D.}~\bibnamefont {Wands}},\
  }\bibfield  {title} {\bibinfo {title} {{Cosmological perturbations through a
  simple bounce}},\ }\href {https://doi.org/10.1103/PhysRevD.70.063515}
  {\bibfield  {journal} {\bibinfo  {journal} {Phys. Rev. D}\ }\textbf {\bibinfo
  {volume} {70}},\ \bibinfo {pages} {063515} (\bibinfo {year}
  {2004})}\BibitemShut {NoStop}%
\bibitem [{\citenamefont {Cai}\ \emph {et~al.}(2011)\citenamefont {Cai},
  \citenamefont {Chen}, \citenamefont {Dent}, \citenamefont {Dutta},\ and\
  \citenamefont {Saridakis}}]{Cai:2011MatterBounce}%
  \BibitemOpen
  \bibfield  {author} {\bibinfo {author} {\bibfnamefont {Y.-F.}\ \bibnamefont
  {Cai}}, \bibinfo {author} {\bibfnamefont {S.-H.}\ \bibnamefont {Chen}},
  \bibinfo {author} {\bibfnamefont {J.~B.}\ \bibnamefont {Dent}}, \bibinfo
  {author} {\bibfnamefont {S.}~\bibnamefont {Dutta}},\ and\ \bibinfo {author}
  {\bibfnamefont {E.~N.}\ \bibnamefont {Saridakis}},\ }\bibfield  {title}
  {\bibinfo {title} {{Matter Bounce Cosmology with the f(T) Gravity}},\ }\href
  {https://doi.org/10.1088/0264-9381/28/21/215011} {\bibfield  {journal}
  {\bibinfo  {journal} {Class. Quant. Grav.}\ }\textbf {\bibinfo {volume}
  {28}},\ \bibinfo {pages} {215011} (\bibinfo {year} {2011})}\BibitemShut
  {NoStop}%
\bibitem [{\citenamefont {Cai}\ \emph {et~al.}(2012)\citenamefont {Cai},
  \citenamefont {Easson},\ and\ \citenamefont
  {Brandenberger}}]{Cai:2012Towards}%
  \BibitemOpen
  \bibfield  {author} {\bibinfo {author} {\bibfnamefont {Y.-F.}\ \bibnamefont
  {Cai}}, \bibinfo {author} {\bibfnamefont {D.~A.}\ \bibnamefont {Easson}},\
  and\ \bibinfo {author} {\bibfnamefont {R.}~\bibnamefont {Brandenberger}},\
  }\bibfield  {title} {\bibinfo {title} {{Towards a Nonsingular Bouncing
  Cosmology}},\ }\href {https://doi.org/10.1088/1475-7516/2012/08/020}
  {\bibfield  {journal} {\bibinfo  {journal} {JCAP}\ }\textbf {\bibinfo
  {volume} {08}},\ \bibinfo {pages} {020}},\ \Eprint
  {https://arxiv.org/abs/1206.2382} {arXiv:1206.2382 [hep-th]} \BibitemShut
  {NoStop}%
\bibitem [{\citenamefont {Nojiri}\ \emph {et~al.}(2019)\citenamefont {Nojiri},
  \citenamefont {Odintsov}, \citenamefont {Oikonomou},\ and\ \citenamefont
  {Paul}}]{Nojiri:2019NonsingularBounce}%
  \BibitemOpen
  \bibfield  {author} {\bibinfo {author} {\bibfnamefont {S.}~\bibnamefont
  {Nojiri}}, \bibinfo {author} {\bibfnamefont {S.~D.}\ \bibnamefont
  {Odintsov}}, \bibinfo {author} {\bibfnamefont {V.~K.}\ \bibnamefont
  {Oikonomou}},\ and\ \bibinfo {author} {\bibfnamefont {T.}~\bibnamefont
  {Paul}},\ }\bibfield  {title} {\bibinfo {title} {{Nonsingular bounce
  cosmology from Lagrange multiplier $F(R)$ gravity}},\ }\href
  {https://doi.org/10.1103/PhysRevD.100.084056} {\bibfield  {journal} {\bibinfo
   {journal} {Phys. Rev. D}\ }\textbf {\bibinfo {volume} {100}},\ \bibinfo
  {pages} {084056} (\bibinfo {year} {2019})}\BibitemShut {NoStop}%
\bibitem [{\citenamefont {Odintsov}\ \emph {et~al.}(2020)\citenamefont
  {Odintsov}, \citenamefont {Oikonomou},\ and\ \citenamefont
  {Paul}}]{Odintsov:2020FromBounce}%
  \BibitemOpen
  \bibfield  {author} {\bibinfo {author} {\bibfnamefont {S.~D.}\ \bibnamefont
  {Odintsov}}, \bibinfo {author} {\bibfnamefont {V.~K.}\ \bibnamefont
  {Oikonomou}},\ and\ \bibinfo {author} {\bibfnamefont {T.}~\bibnamefont
  {Paul}},\ }\bibfield  {title} {\bibinfo {title} {{From a Bounce to the Dark
  Energy Era with $F(R)$ Gravity}},\ }\href
  {https://doi.org/10.1088/1361-6382/abbc47} {\bibfield  {journal} {\bibinfo
  {journal} {Class. Quant. Grav.}\ }\textbf {\bibinfo {volume} {37}},\ \bibinfo
  {pages} {235005} (\bibinfo {year} {2020})}\BibitemShut {NoStop}%
\bibitem [{\citenamefont {Nojiri}\ \emph {et~al.}(2005)\citenamefont {Nojiri},
  \citenamefont {Odintsov},\ and\ \citenamefont
  {Tsujikawa}}]{nojiri2005properties}%
  \BibitemOpen
  \bibfield  {author} {\bibinfo {author} {\bibfnamefont {S.}~\bibnamefont
  {Nojiri}}, \bibinfo {author} {\bibfnamefont {S.~D.}\ \bibnamefont
  {Odintsov}},\ and\ \bibinfo {author} {\bibfnamefont {S.}~\bibnamefont
  {Tsujikawa}},\ }\bibfield  {title} {\bibinfo {title} {{Properties of
  singularities in (phantom) dark energy universe}},\ }\href
  {https://doi.org/10.1103/PhysRevD.71.063004} {\bibfield  {journal} {\bibinfo
  {journal} {Phys. Rev. D}\ }\textbf {\bibinfo {volume} {71}},\ \bibinfo
  {pages} {063004} (\bibinfo {year} {2005})}\BibitemShut {NoStop}%
\bibitem [{\citenamefont {de~Haro}\ \emph {et~al.}(2023)\citenamefont
  {de~Haro}, \citenamefont {Nojiri}, \citenamefont {Odintsov}, \citenamefont
  {Oikonomou},\ and\ \citenamefont {Pan}}]{deHaro:2023Finite-time}%
  \BibitemOpen
  \bibfield  {author} {\bibinfo {author} {\bibfnamefont {J.}~\bibnamefont
  {de~Haro}}, \bibinfo {author} {\bibfnamefont {S.}~\bibnamefont {Nojiri}},
  \bibinfo {author} {\bibfnamefont {S.~D.}\ \bibnamefont {Odintsov}}, \bibinfo
  {author} {\bibfnamefont {V.~K.}\ \bibnamefont {Oikonomou}},\ and\ \bibinfo
  {author} {\bibfnamefont {S.}~\bibnamefont {Pan}},\ }\bibfield  {title}
  {\bibinfo {title} {{Finite-time cosmological singularities and the possible
  fate of the Universe}},\ }\href
  {https://doi.org/10.1016/j.physrep.2023.09.003} {\bibfield  {journal}
  {\bibinfo  {journal} {Phys. Rept.}\ }\textbf {\bibinfo {volume} {1034}},\
  \bibinfo {pages} {1} (\bibinfo {year} {2023})}\BibitemShut {NoStop}%
\bibitem [{\citenamefont {Katsuragawa}\ \emph {et~al.}(2024)\citenamefont
  {Katsuragawa}, \citenamefont {Nojiri},\ and\ \citenamefont
  {Odintsov}}]{Katsuragawa:2024FutureSingularity}%
  \BibitemOpen
  \bibfield  {author} {\bibinfo {author} {\bibfnamefont {T.}~\bibnamefont
  {Katsuragawa}}, \bibinfo {author} {\bibfnamefont {S.}~\bibnamefont
  {Nojiri}},\ and\ \bibinfo {author} {\bibfnamefont {S.~D.}\ \bibnamefont
  {Odintsov}},\ }\bibfield  {title} {\bibinfo {title} {{Future singularity in
  anisotropic universe}},\ }\href@noop {} {\bibfield  {journal} {\bibinfo
  {journal} {arXiv preprint arXiv:2406.18368}\ } (\bibinfo {year} {2024})},\
  \Eprint {https://arxiv.org/abs/2406.18368} {arXiv:2406.18368 [gr-qc]}
  \BibitemShut {NoStop}%
\bibitem [{\citenamefont {Bernal}\ \emph {et~al.}(2016)\citenamefont {Bernal},
  \citenamefont {Verde},\ and\ \citenamefont
  {Riess}}]{Bernal:2016TheTroubleH0}%
  \BibitemOpen
  \bibfield  {author} {\bibinfo {author} {\bibfnamefont {J.~L.}\ \bibnamefont
  {Bernal}}, \bibinfo {author} {\bibfnamefont {L.}~\bibnamefont {Verde}},\ and\
  \bibinfo {author} {\bibfnamefont {A.~G.}\ \bibnamefont {Riess}},\ }\bibfield
  {title} {\bibinfo {title} {{The trouble with $H_0$}},\ }\href
  {https://doi.org/10.1088/1475-7516/2016/10/019} {\bibfield  {journal}
  {\bibinfo  {journal} {JCAP}\ }\textbf {\bibinfo {volume} {10}},\ \bibinfo
  {pages} {019}}\BibitemShut {NoStop}%
\bibitem [{\citenamefont {Vagnozzi}(2023)}]{Vagnozzi:2023Seven}%
  \BibitemOpen
  \bibfield  {author} {\bibinfo {author} {\bibfnamefont {S.}~\bibnamefont
  {Vagnozzi}},\ }\bibfield  {title} {\bibinfo {title} {{Seven Hints That
  Early-Time New Physics Alone Is Not Sufficient to Solve the Hubble
  Tension}},\ }\href {https://doi.org/10.3390/universe9090393} {\bibfield
  {journal} {\bibinfo  {journal} {Universe}\ }\textbf {\bibinfo {volume} {9}},\
  \bibinfo {pages} {393} (\bibinfo {year} {2023})}\BibitemShut {NoStop}%
\end{thebibliography}%

\end{document}